\documentclass[twocolumn,tighten]{aastex63}

\received{\today}
\revised{\today}
\accepted{\today}

\usepackage{bm}
\usepackage{amsmath}
\usepackage{hyperref}

\makeatletter
\newcommand*\bigcdot{\mathpalette\bigcdot@{.5}}
\newcommand*\bigcdot@[2]{\mathbin{\vcenter{\hbox{\scalebox{#2}{$\m@th#1\bullet$}}}}}
\makeatother

\shorttitle{Particles \& Self-Gravity II: Accretion and Dust}
\shortauthors{Baehr \& Zhu}
\graphicspath{{./}{figures/}}

\begin{document}

\title{Particle Dynamics in 3D Self-gravitating Disks II:  Strong Gas Accretion and Thin Dust Disks}  

\correspondingauthor{Hans Baehr}
\email{hans-paul.baehr@unlv.edu}

\author[0000-0002-0880-8296]{Hans Baehr}
\affil{Department of Physics and Astronomy, University of Nevada, Las Vegas, 4505 South Maryland Parkway, Las Vegas, NV 89154, USA}

\author[0000-0003-3616-6822]{Zhaohuan Zhu}
\affil{Department of Physics and Astronomy, University of Nevada, Las Vegas, 4505 South Maryland Parkway, Las Vegas, NV 89154, USA}




\begin{abstract}
Observations suggest that protoplanetary disks have moderate accretion rates onto the central young star, especially at early stages (e.g. HL Tau), indicating moderate disk turbulence. However, recent ALMA observations suggest that dust is highly settled, implying weak turbulence. Motivated by such tension, we carry out 3D stratified local simulations of self-gravitating disks, focusing on settling of dust particles in actively accreting disks. We find that gravitationally unstable disks can have moderately high accretion rates while maintaining a relatively thin dust disk for two reasons. First, accretion stress from the self-gravitating spirals (self-gravity stress) can be stronger than the stress from turbulence (Reynolds stress) by a factor of 5-20. Second, the strong gravity from the gas to the dust decreases the dust scale height by another factor of $\sim 2$. Furthermore, the turbulence is slightly anisotropic, producing a larger Reynolds stress than the vertical dust diffusion coefficient. Thus, gravitoturbulent disks have unusually high vertical Schmidt numbers ($Sc_z$) if we scale the total accretion stress with the vertical diffusion coefficient (e.g. $Sc_z\sim$ 10-100). The reduction of the dust scale height by the gas gravity, should also operate in gravitationally stable disks ($Q>$1). Gravitational forces between particles become more relevant for the concentration of intermediate dust sizes, forming dense clouds of dust. After comparing with HL Tau observations, our results suggest that self-gravity and gravity among different disk components could be crucial for solving the conflict between the protoplanetary disk accretion and dust settling, at least at the early stages.
\end{abstract}

\keywords{protoplanetary disks --- turbulence --- hydrodynamics}

\section{Introduction}
\label{sec:intro}

Disks accrete significant amounts of material during their $\sim$ 1 Myr lifetime \citep{Ribas2015}, but the process which can ubiquitously transfer angular momentum in a disk remains elusive. Measurements of gas turbulence often cannot reveal the midplane properties, due to optically thick tracers. Normally only gas velocities in the upper disk atmosphere can be probed \citep{Flaherty2015,Teague2016,Flaherty2017}. Observing the dust layer allows for a different indicator of disk turbulence at the midplane \citep{Pinte2016}.

Various fluid instabilities have been used to explain the turbulence which can drive disk accretion, including, but not limited to, the magneto-rotational instability (MRI) \citep{Balbus1991}, subcritical baroclinic instability (SBI) \citep{Klahr2003,Klahr2014}, and vertical shear instability (VSI) \citep{Nelson2013}. Each is promising but comes with important caveats. SBI requires low viscosities and produces comparatively weak angular momentum transport \citep{Malygin2017,Pfeil2019} while MRI appears less likely to operate in the midplane of protostellar disks due to low ionization and non-ideal MHD effects \citep{Lesur2014,Bai2015}.

VSI has become a focal point of research to explain disk turbulence due to its robustness and being capable of producing other disk features (e.g. vortices) \citep{Stoll2014,Manger2018}. However, \citet{Flock2020} recently showed that for the level of turbulent stress that VSI produces, the measured dust distribution is thicker than in observations. \cite{Pinte2016} has shown that the dust disk of HL Tau revealed by ALMA has a scale height less than 1 AU at 100 AU, which is equivalent to a turbulent level of $\alpha<10^{-3}$. On the other hand, the detected accretion rate requires a turbulent level of $\alpha\sim10^{-2}$. Furthermore, recent edge-on ALMA observations of dust disks \citep{Villenave2020} and disk accretion rate constraints from spectral energy distribution studies \citep{Ribas2020} add more evidence that dust layers are significantly thinner than theoretically anticipated. This suggests particles are significantly settled even though there is still moderate accretion to drive short disk lifetimes.

In this paper, we focus on the vertical dust distribution in disks subject to gravitational instabilities (GI) which make it an alternative to hydrodynamic instabilities like VSI. We lay out the theoretical framework on GI and dust diffusion in Section \ref{sec:gravitoturbulence}. Section \ref{sec:model} describes our numerical setup. In section \ref{sec:results}, we measure the particle scale heights and gas turbulent properties to determine the level of vertical dust settling and vertical diffusion. Finally, the implications on disk observations and the evolution of protoplanetary disks are discussed in Section \ref{sec:discussion} before we summarize our findings in Section \ref{sec:conclusion}.

\begin{figure*}[t]
\centering
\includegraphics[width=0.49\textwidth]{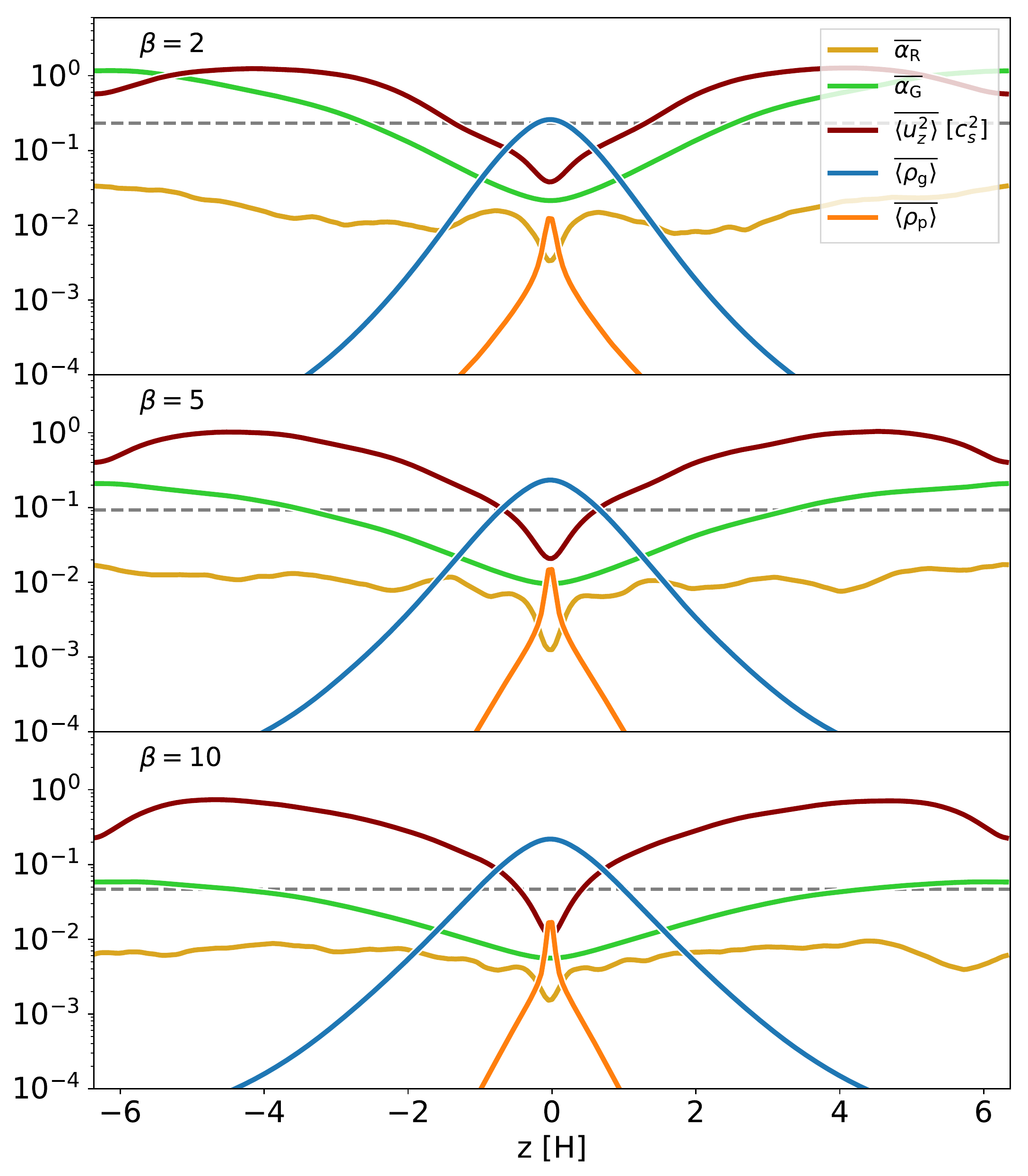}%
\includegraphics[width=0.49\textwidth]{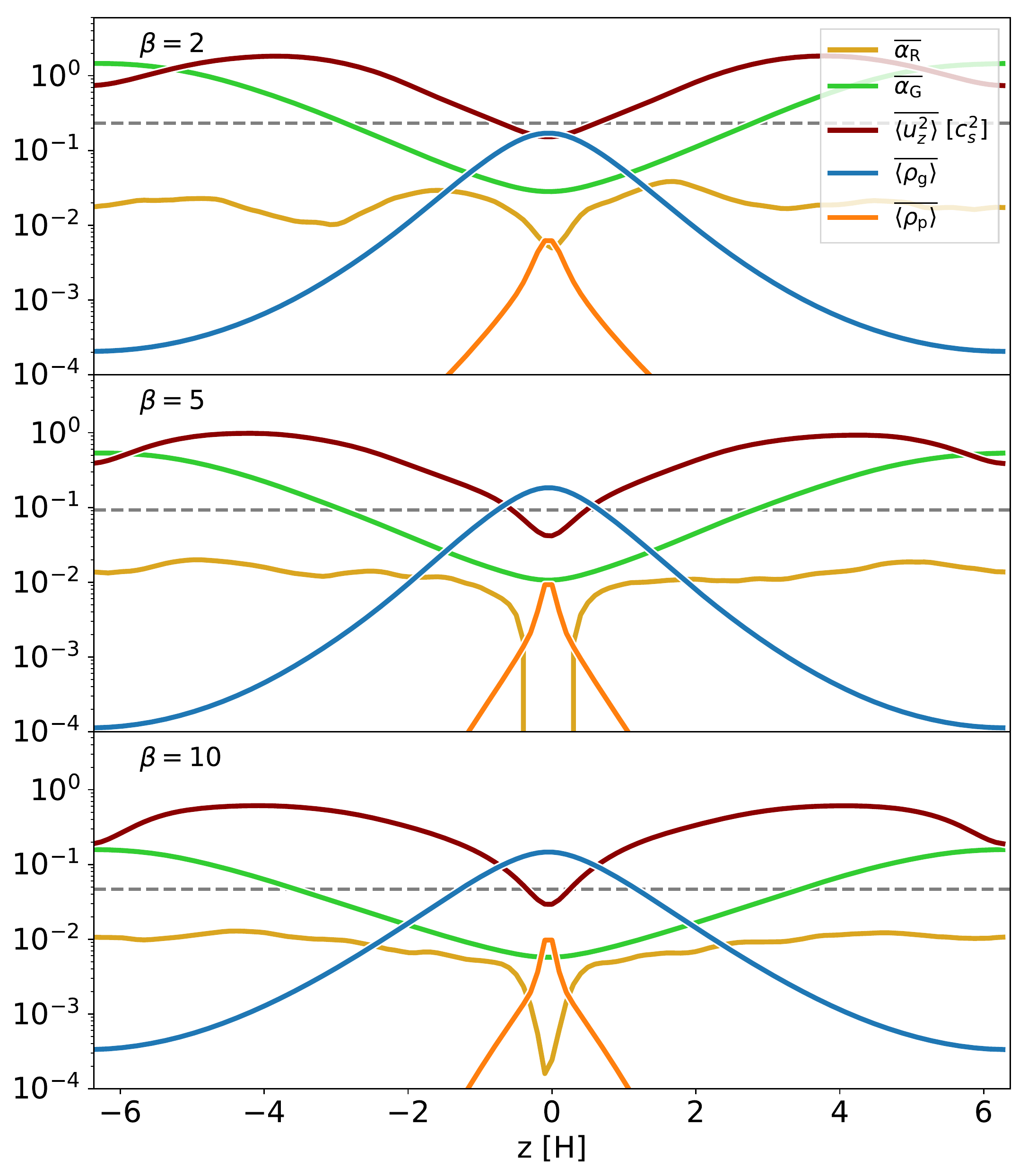}
\caption{Time-averaged density and turbulent vertical profiles for each cooling timescale $\beta$. Left panel: medium-sized simulations ($L_{x} = L_{y} = 25H$) Right panel: large-sized simulations ($L_{x} = L_{y} = 51H$). Solid lines indicate the gas density (blue), dust density (orange), vertical gas velocity squared (dark red), Reynolds stress (gold) and gravitational stress (green). The gray dashed line is the total $\alpha$ from Equation \eqref{eq:gammiealpha}.}
\label{fig:densityturbulenceprofile}
\end{figure*}

\section{Gravitoturbulence}
\label{sec:gravitoturbulence}

Turbulent transport in disks is described via the $\alpha$ model \citep{Shakura1973}, where the viscosity $\nu$ is scaled to the local isotropic pressure
\begin{equation} \label{eq:alphaviscosity}
\nu = \alpha c_{\mathrm{s}} H\,,
\end{equation}
such that $\alpha$ is a dimensionless constant describing the relative size and speed of turbulent motions in terms of the disk pressure scale height $H=c_s/\Omega$ and the sound speed $c_{s}$ \citep{Pringle1981}. Please note that we define $c_s=\sqrt{\gamma P/\rho}$, which has also been adopted for defining the Toomre Q parameter and disk $\alpha$ \citep{Gammie2001}. In a thin disk approximation, the remaining component of the vertically-integrated stress tensor $\bm{T}$ can be expressed in terms of $\alpha$, pressure $P$, orbital radius $R$ and orbital frequency $\Omega$ \citep{Cossins2009}
\begin{equation} \label{eq:stresstensor}
T_{r\phi} = -\alpha P \frac{d\ln \Omega}{d \ln R}.
\end{equation}
The stress tensor can be separated into two components,  the gravitational stress $G$ (e.g. due to spiral's self-gravity) and the Reynolds stress $R$ from the disk turbulence, 
\begin{equation}
R + G = -\alpha P \frac{d\ln \Omega}{d \ln R}\,,
\end{equation}
where $d\ln \Omega/d \ln R=-3/2$ in a Keplerian disk.
Thus, one can describe the total $\alpha$ stress 
\begin{equation} \label{eq:totalalphastress}
\alpha = \alpha_{R} + \alpha_{G}.
\end{equation}
as the sum of the Reynolds alpha
\begin{equation}\label{eq:reynoldsstress}
\alpha_{R} = \frac{2}{3} \frac{\langle \rho u_{x}u_{y} \rangle}{\langle \rho c_{\mathrm{s}}^{2} \rangle},
\end{equation}
and the gravitational alpha
\begin{equation}
\alpha_{G} = \frac{2}{3} \frac{\langle g_{x}g_{y} \rangle}{4\pi G\langle \rho c_{\mathrm{s}}^{2} \rangle},
\end{equation}
where $g_{x}$ and $g_{y}$ are the local gravitational accelerations in the $x$ and $y$ directions respectively and $P=\rho c_{s}^{2}$ is the locally isothermal pressure. In this paper, angled brackets are the average over the horizontal $x-y$ plane for each vertical position. 

Circumstellar disks are susceptible to non-axisymmetric density waves and gravitoturbulence as long as $1 < Q \lesssim 1.5$ \citep{Durisen2007} where
\begin{equation}\label{eq:toomre}
Q = \frac{c_{s}\Omega}{\pi G\Sigma}\,,
\end{equation}
is the Toomre stability parameter \citep{Toomre1964}. Disks will fragment when the both the thermal relaxation time $t_{c} = \beta\Omega^{-1} \lesssim 3$ and when $Q<1$ \citep{Gammie2001}. To avoid fragmentation, we consider $\Omega =G=1$, $c_{s,0} = \pi$ and the combined gas and dust surface density $\Sigma_{\mathrm{0}} = \Sigma_{\mathrm{g,0}} + \Sigma_{\mathrm{p,0}}$ such that $Q(t=0)$ is just slightly above unity.

Depending on the cooling efficiency $\beta$, gravitoturbulence can be quite strong, i.e. $\alpha \sim 0.01-0.1$   (\citealt{Gammie2001},but see \citealt{Paardekooper2012}). However, without mass replenishment, the turbulence can die out quickly. The accretion led by the instability  decreases the mass which in turn makes the disk stable against the instability, and only with enough accretion from the envelope onto the disk can gravitational instabilities persist \citep{Zhu2012}.

Recent studies that have looked into the vertical structure of gravitoturbulence find that turbulent stresses and turbulent velocities vary only by an order of magnitude from the midplane to a few scale heights despite much steeper decreases in gas density \citep{Shi2014,Riols2017}. It is important to distinguish between the stress components $R$ and $G$ and their dimensionless counterparts, $\alpha_{R}$ and $\alpha_{G}$. In stratified simulations, the vertical density profile and thus the vertical pressure profile will lead to separate structure for both components. The initial temperature profile is isothermal and it remains vertically isothermal when in equilibrium, so the density and pressure profiles scale the same with distance from the midplane. Reynolds stresses scale similar to the pressure, leading to a mostly flat profile of $\alpha_{R}$. Gravitational stresses $G$ are stronger near the midplane but only drop off modestly with increasing height above the midplane, highlighting the range and strength of gravitational interactions \citep{Shi2014}. Since the pressure drops off rapidly with height above the midplane compared to the gravitational stress, $\alpha_G$ will increase significantly even though the gravitational stress is decreasing \citep{Shi2014}. Thus $\alpha_G$ is comparable to $\alpha_R$ near the midplane ($\sim 10^{-2}$), but in some cases approaching order unity in the upper levels of the disk atmosphere. This is potentially relevant to observations, as only a fraction of disk accretion is due to the turbulence which can diffuse the dust.

More recently, \citet{Riols2020} investigated the vertical structure of both the gas and dust in gravitoturbulent disks in order to ascertain how well turbulence models compare to well-coupled dust measured at and around the midplane in observations. Whether the long range vertical effects of self-gravity in comparison to the local midplane turbulence can affect the settling of dust merits closer attention.

Besides the fact that the total $\alpha$ can be significantly larger than the turbulent $\alpha_R$ which can lead to a thin dust disk, the vertical gravity from gas to dust can make the dust disk even thinner. We can derive the density structure of the dust disk by equating the settling flux with the diffusion flux. For small particles, the terminal velocity can be calculated by balancing the gravitational force with the gas drag force
\begin{equation}
-(F_{sg}(z)+\Omega^2 z)=\frac{u_{z}}{t_{s}}\,,
\end{equation}
where $F_{sg}(z)$ is the vertical force due to disk self-gravity. If we assume that the disk structure is uniform in the radial and azimuthal directions at the scale of the disk scale height, we can use Gauss' theorem to derive the self-gravity force, so that
\begin{equation}
-(2\pi G \Sigma(z)+\Omega^2 z)=\frac{u_{z}}{t_{s}}\,,
\end{equation}
where $\Sigma(z)$ is the gas surface density, including both gas and dust, integrated from the midplane to the height of $z$. Then the flux balance is
\begin{equation}
-(2\pi G \Sigma(z)+\Omega^2 z)t_{s}\rho_{d}=D_{d,z}\rho_{g}\frac{d(\rho_{d}/\rho_{g})}{dz}\,.
\end{equation}
This is an integro-differential equation since  $\Sigma$ is the integral of $\rho_d+\rho_g$. If we assume that most mass is concentrated at the midplane, we can replace $\Sigma(z)$ with $\Sigma$ of the disk. And if we assume that the gas stratification is much less than the dust stratification, $\rho_{g}$ can be canceled on the right side of the equation. Then, we can derive
\begin{equation}
\frac{\rho_{d}}{\rho_{d,mid}}=e^{-\frac{t_{s}}{D_{d,z}}\left(2\pi G\Sigma z + \frac{\Omega^2 z^2}{2} \right)}\,.
\end{equation}
When self-gravity is ignored, this returns to the known dust scale height of
$H_{d,0}=\sqrt{D_{d,z}/(\Omega^2 t_s)}$. If we use $H_{d,0}$ in the equation we can derive
\begin{equation}
\frac{\rho_{d}}{\rho_{d,mid}}=e^{-\frac{2 H z}{Q H_{d,0}^2  }-\frac{z^2}{2 H_{d,0}^2}}\,.
\end{equation}
Thus, at $z=H_{d,0}$, the effect of self gravity (the first term in the exponential) is stronger than the gravity from the star (the second term in the exponential) by a factor of $4H/(Q H_{d,0})$. Although, this factor can become quite large if $H\gg H_{d,0}$, our derivation only applies if most of the disk mass is at the midplane. Considering that the gas mass is far larger than the dust mass, it means that our derivation applies to dust that is not very settled (e.g. $H_{d,0}\sim H$).

\begin{figure}
\centering
\includegraphics[width=0.48\textwidth]{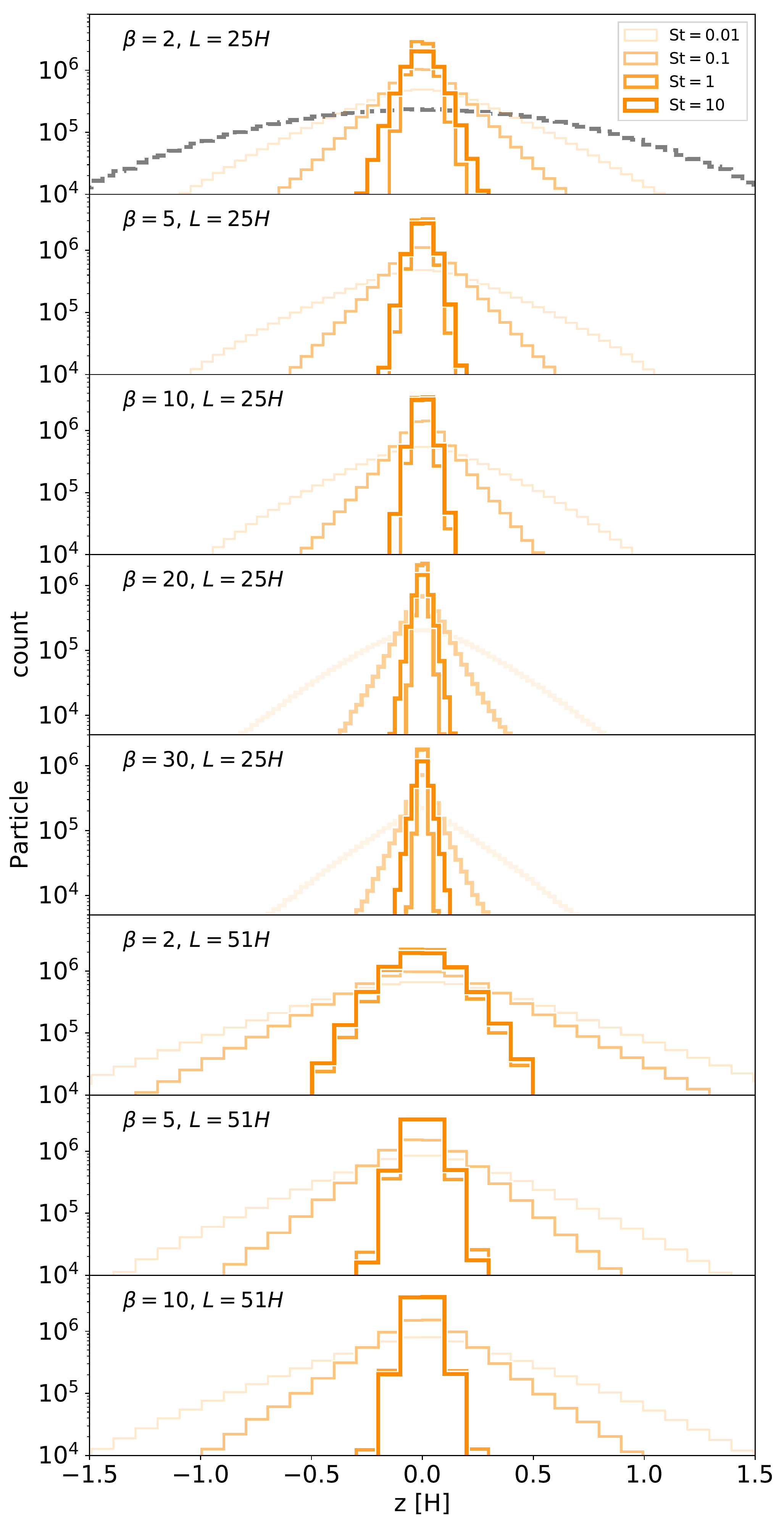}
\caption{Cumulative number of particles in the time interval $t=50\,\Omega^{-1}$ to $t=80\,\Omega^{-1}$. For each simulation, dust particles have been separated into each species and binned to the nearest vertical grid cell. The initial distribution is plotted as a gray dashed line in the top frame for comparison, adjusted for the number of snapshots.}
\label{fig:agglomMultispecies}
\end{figure}

\begin{figure}
\centering
\includegraphics[width=0.48\textwidth]{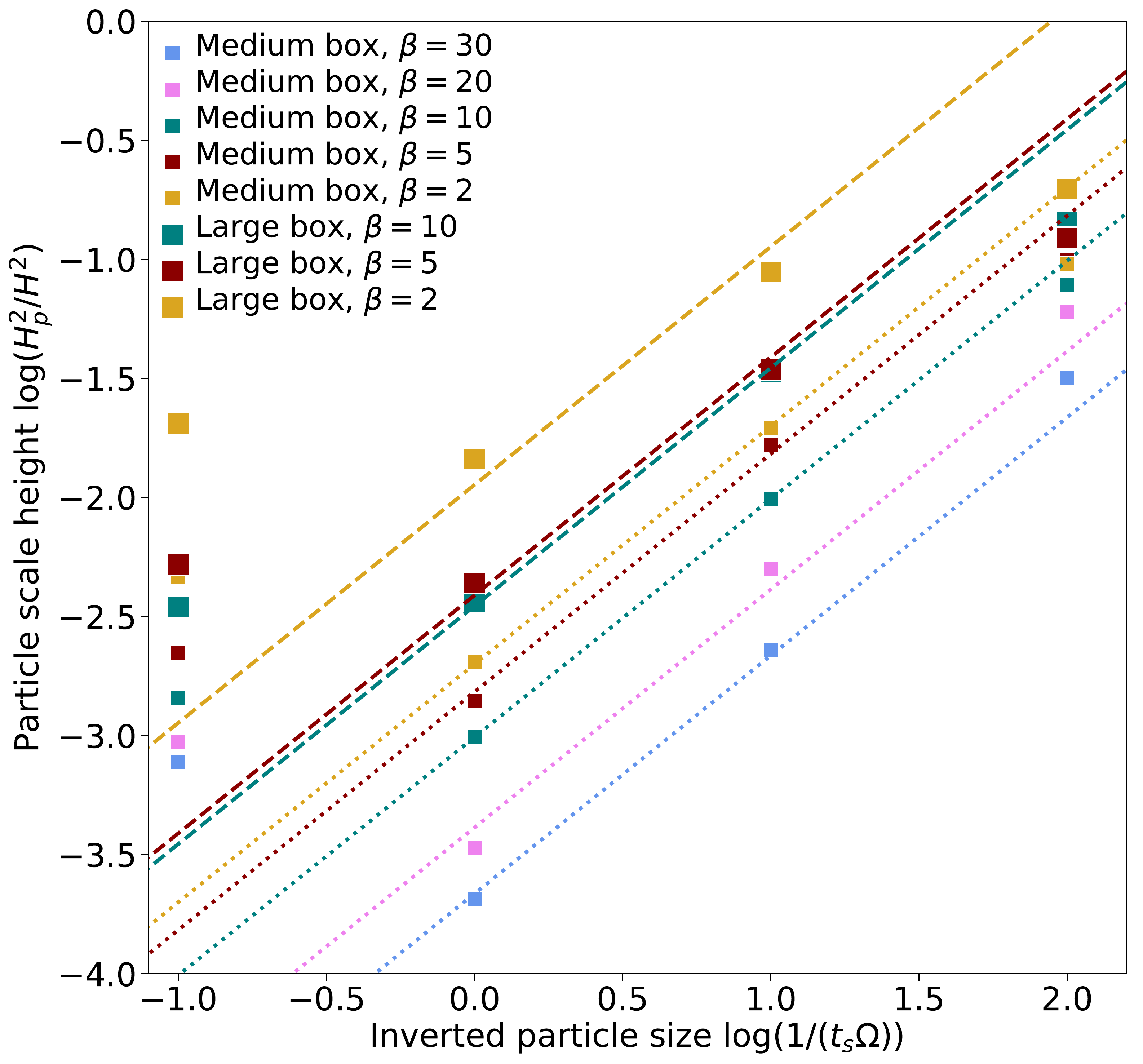}
\caption{Particle scale height $H_{\mathrm{p}}^{2}$ as a function of the inverted particle size $t_{\mathrm{s}}^{-1} = \Omega\mathrm{St}^{-1}$ for the medium boxes (smaller squares, fit with dotted lines) and large boxes (larger squares, fit with dashed lines). Fitting a straight line to the intermediate sized particles $(\mathrm{St}=0.1,1)$ results in a slope equal to the diffusion parameter $D_{d,z}$ (Equation \eqref{eq:diffusionparameter}), which we estimate for both medium and large setups separately.}
\label{fig:diffusionparameter}
\end{figure}

On the other hand, for highly settled dust, we can assume that the self-gravity from gas dominates and the gas is almost uniform close to the disk midplane. Then, we can replace $\Sigma(z)$ with $\rho_{mid}z$, and the vertical density structure is
\begin{equation}
\frac{\rho_{d}}{\rho_{d,mid}}=e^{-\frac{t_{s}}{D_{d,z}}\left(\frac{(\Omega^2+2\pi G\rho_{mid}) z^2}{2}\right)}\,.
\end{equation}
If we use $\rho_{\mathrm{mid}}\sim\Sigma/(2H_{0})$, the equation becomes
\begin{equation}
\frac{\rho_{d}}{\rho_{d,mid}}=e^{-\frac{t_{s}}{D_{d,z}}\left(\frac{\Omega^2(1+ 1/Q) z^2}{2}\right)}\,.
\end{equation}
Thus, the dust scale height is 
\begin{equation}
H_{d}=\sqrt{\frac{Q}{Q+1}\frac{D_{\mathrm{d},z}}{\Omega^2 t_{s}}}    \,.\label{eq:selfgravitydust}
\end{equation}
When $Q\sim 1$, the dust scale height is 71\% of the dust scale height without considering the disk self-gravity. Thus, the self-gravity of the gas alone could suppress vertical diffusion of small dust particles and increase settling.

\begin{deluxetable*}{ccccccccccc}
\tablecaption{Turbulent properties for each simulation}

\tablehead{\colhead{Simulation} & \colhead{$\overline{Q}$}  & \colhead{$\overline{\alpha}$} & \colhead{$\overline{\alpha_{R}}$} & \colhead{$\overline{\alpha_{G}}$} & \colhead{$\overline{u^{2}_{z}}$ $[c_{\mathrm{s}}^{2}]$} & \colhead{$\overline{u^{2}_{z,\rho}}$ $[c_{\mathrm{s}}^{2}]$} & \colhead{$\overline{H_{0.1}/H_{g}}$} & \colhead{$D^{'}_{\mathrm{d,z}}$} & \colhead{$\mathrm{Sc}_{z}$} & \colhead{$\mathrm{Sc}_{\mathrm{R},z}$}}
\startdata
S\_t2\_B  & $1.2$ & $4.9\times 10^{-1}$ & $1.6\times 10^{-2}$ & $4.7\times 10^{-1}$ & $0.74$ & $0.64$ & $0.32$  & $2.0\times 10^{-3}$   & $245$ & $8$\\
S\_t5\_B  & $1.3$ & $1.0\times 10^{-1}$ & $1.0\times 10^{-2}$ & $9.4\times 10^{-2}$ & $0.56$ & $0.51$ & $0.28$  & $1.5\times 10^{-3}$   & $67$  & $6.6$\\
S\_t10\_B & $1.3$ & $3.9\times 10^{-2}$ & $6.3\times 10^{-3}$ & $3.2\times 10^{-2}$ & $0.40$ & $0.37$ & $0.25$  & $1.0\times 10^{-3}$   & $39$  & $6$\\
S\_t2\_BB & $1.5$ & $5.6\times 10^{-1}$ & $2.0\times 10^{-2}$ & $5.4\times 10^{-1}$ & $1.04$ & $0.86$ & $0.22$  & $1.1\times 10^{-2}$   & $43$  & $1.8$\\
S\_t5\_BB & $1.5$ & $2.1\times 10^{-1}$ & $1.2\times 10^{-2}$ & $2.0\times 10^{-1}$ & $0.56$ & $0.48$ & $0.22$  & $3.9\times 10^{-3}$   & $53$ & $3.1$ \\
S\_t10\_BB & $1.7$ & $6.7\times 10^{-2}$ & $8.5\times 10^{-3}$ & $5.9\times 10^{-2}$ & $0.37$ & $0.33$ & $0.18$  & $3.5\times 10^{-3}$   & $19$ & $2.4$\\
\hline
S\_t5\_B\_lowpsg & $1.2$ & $9.3\times 10^{-2}$ & $9.6\times 10^{-3}$ & $8.3\times 10^{-2}$ & $0.55$ & $0.49$ & $0.31$  & $1.3\times 10^{-3}$  & $71$ & $7.4$  \\
S\_t5\_B\_nopsg & $1.3$ & $1.1\times 10^{-1}$ & $9.4\times 10^{-3}$ & $1.0\times 10^{-1}$ & $0.54$ & $0.46$ & $0.30$  & $5.3\times 10^{-3}$  & $20$ & $1.7$ \\
S\_t5\_B\_nogi  & - & - & - & - & 0.07 & 0.07 & $0.20$  & -  & - & -\\
\hline
S\_t20\_B\_hires & $1.3$ & $2.2\times 10^{-2}$ & $4.8\times 10^{-3}$ & $1.7\times 10^{-2}$ & $0.35$ & $0.32$ & $0.26$  & $4\times 10^{-4}$  & $55$ & $12$  \\
S\_t30\_B\_hires & $1.3$ & $1.0\times 10^{-2}$ & $2.4\times 10^{-3}$ & $7.6\times 10^{-3}$ & $0.24$ & $0.22$ & $0.24$  & $2\times 10^{-4}$  & $50$ & $12$ \\
\enddata
\tablecomments{Diagnostics for the turbulent properties of each simulation, including Toomre stability parameter $Q$, turbulent $\alpha$ parameter with Reynolds and gravitational components $\alpha_{\mathrm{R}}$ and $\alpha_{\mathrm{G}}$, vertical gas velocity squared, calculated both as a volume average $u^{2}_{z}$ and a density-weighted average $u^{2}_{z,\rho}$, the ratio of particle scale height to gas scale height at $\mathrm{St}=0.1$ $H_{0.1}/H_{g}$, dimensionless diffusion constant $D^{'}_{\mathrm{d,z}}$, vertical Schmidt number with respect to the total stress $\mathrm{Sc}_{z}$, and vertical Schmidt number with respect to the Reynolds stress only $\mathrm{Sc}_{\mathrm{R},z}$. Quantities with a bar are averaged over 30 snapshots from $t=50\,\Omega^{-1}$ to $t=80\,\Omega^{-1}$ to cancel out some of the noise at individual snapshots.}
\label{tab:turbresults}
\end{deluxetable*}

\section{Model}
\label{sec:model}

As in the first paper \citep{Baehr2021}, we use 3D hydrodynamic shearing box simulations to study the dynamics of self-gravitating disks with Lagrangian super-particles embedded in the Eulerian mesh using the \textsc{Pencil} code \citep{Brandenburg2003}.

We set our simulations to be marginally Toomre stable, $Q_{0} = 1.02$. The simulation details are given in the model section and Table 1 of \citet{Baehr2021}. We briefly summarize the simulations below. Simulations are designated by their cooling timescale and size. For example, the simulation named \texttt{S\_t2\_B} is the simulation with cooling time $\beta = 2$ and a medium-sized domain $L_{x}=L_{y}=(80/\pi) H$, $L_{z}=(40/\pi) H$ with $512^2\times256$ grid cells. The suffix \texttt{BB} refers to big box simulations with $L_{x}=L_{y}=(160/\pi) H$, $L_{z}=(40/\pi) H$ with $512^2\times128$ grid cells. We picked three cooling times $\beta=$ 2, 5, and 10. All simulations include $1.5\times 10^{6}$ particles evenly distributed among the 6 species $\mathrm{St}=[0.01,0.1,1,10,100,1000]$, and the initial dust-to-gas mass ratio is $\epsilon = 0.01$. Simulations with self-gravity of the gas and dust turned on or off are described below. Additional simulations were run at cooling timescales comparable to those in \citet{Riols2020}, and at higher grid resolution.

Gas and dust are both initialized with Gaussian distributions with the same vertical width $H_{\mathrm{g},0}=H_{\mathrm{d},0}=0.63 H$ and allowed to relax as turbulence and settling develop. Stellar gravity is included with a sinusoidal vertical profile to avoid a discontinuity of the vertical acceleration at the periodic upper and lower boundaries. Simulations are either resolved in each direction by 10 grid cells per $H$ for the large boxes or 20 grid cells per $H$ for the medium boxes. We compare these resolutions with other studies in Section \ref{subsec:additionalconsiderations}.

\begin{figure}[t]
\centering
\includegraphics[width=0.48\textwidth]{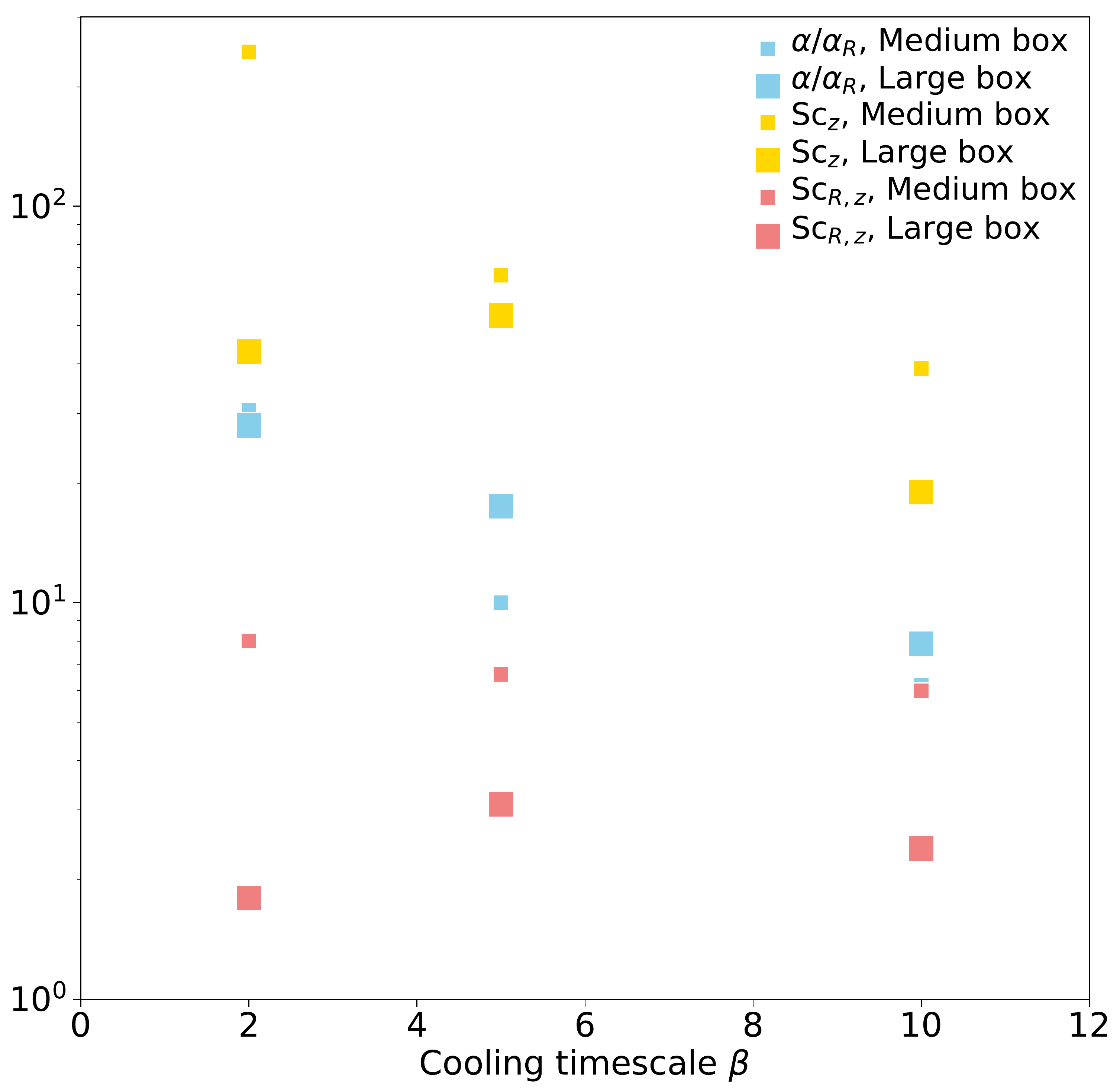}
\caption{Turbulent diagnostics versus the cooling timescale $\beta$. While the Reynolds stress scales similar to the vertical diffusion and $\mathrm{Sc}_{R,z}$ (red squares) is roughly constant, $\alpha/\alpha_{R}$ (blue squares) and $\mathrm{Sc}_{z}$ (yellow squares) both decline as $\beta$ increases and gravitoturbulence becomes weaker.}
\label{fig:turbulentratios}
\end{figure}

\begin{figure*}[t]
\centering
\includegraphics[width=0.48\textwidth]{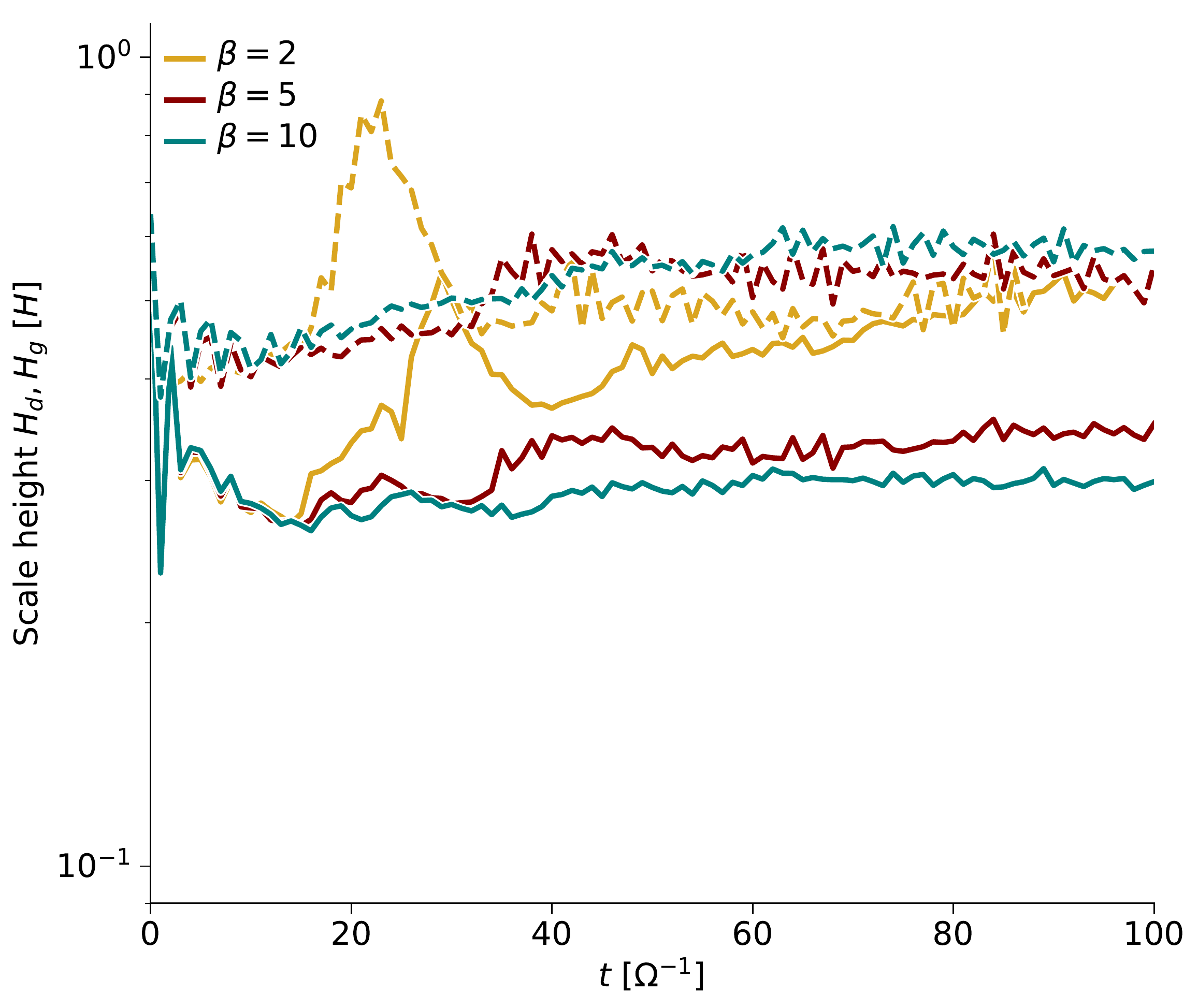}%
\includegraphics[width=0.48\textwidth]{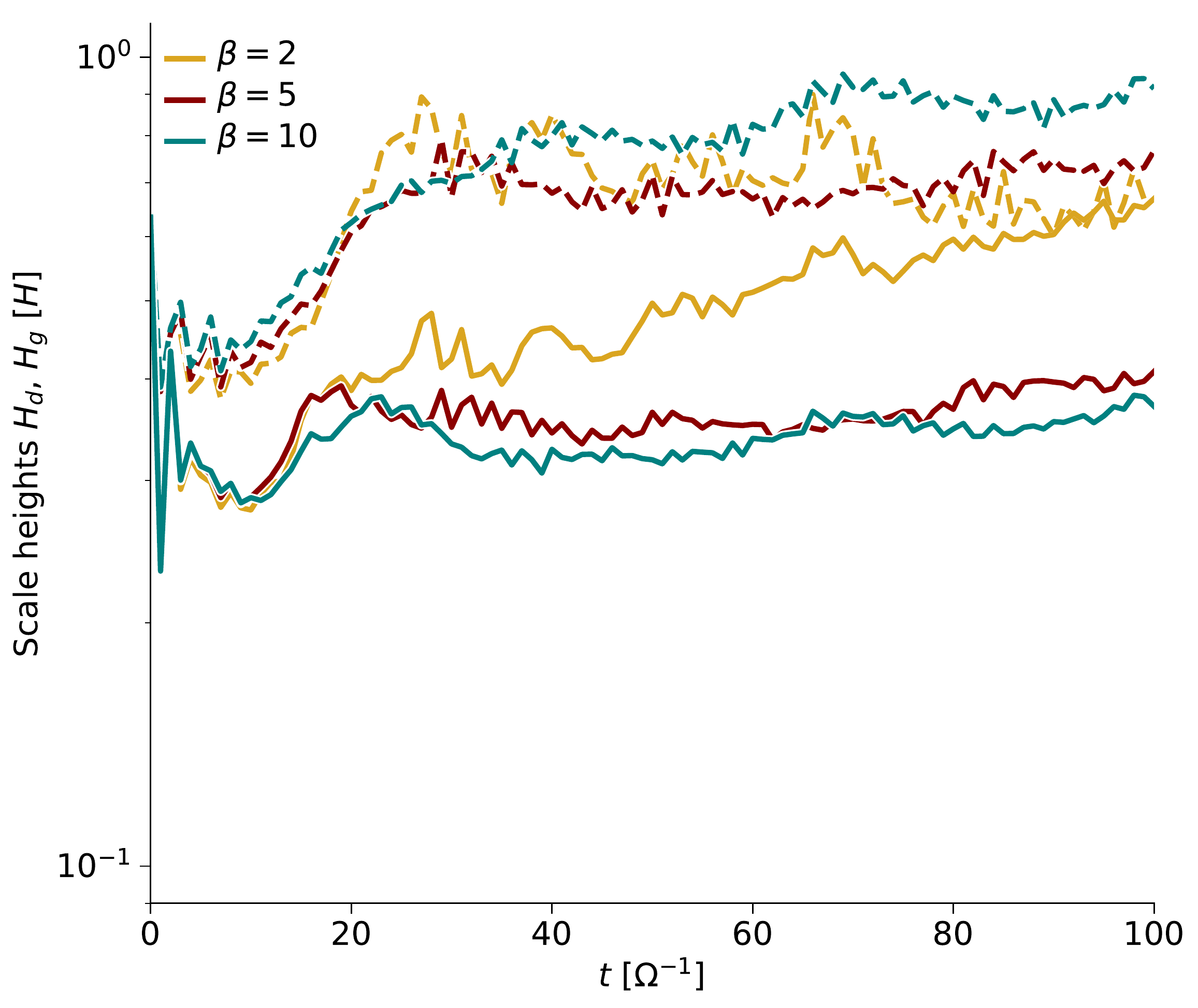}
\caption{Evolution of measured dust ($H_{d}$, solid lines) and gas ($H_{g}$, dashed lines) scale heights over time in our medium (left) and large (right) sized simulations. The dust scale height is measured from the distribution of all dust species combined.}
\label{fig:scaleheightevolution}
\end{figure*}

Self-gravity of the gas and dust is solved in Fourier space by transforming the density to find the potential at wavenumber $k$ and transforming the solution back into real space. The solution to the Poisson equation in Fourier space at wavenumber $\bm{k} = (k_{x},k_{y},k_{z})$ is
\begin{equation} \label{eq:gravpotential}
\Phi(\bm{k}, t) = -\frac{2\pi G\rho(\bm{k}, t)}{\bm{k}^2},
\end{equation}
where $\Phi = \Phi_{\mathrm{g}} + \Phi_{\mathrm{d}}$ and $\rho = \rho_{\mathrm{g}} + \rho_{\mathrm{d}}$ are the potential and density of the gas plus dust particles combined. This is the default setup for the first six simulations and we distinguish these six from the next three by the inclusion of gas-particle and particle-particle gravitational interactions self-consistently. Among these three simulations, the simulation with 'low dust mass' (simulation suffix \texttt{lowpsg}) indicates that the particle mass is so low that it does not contribute to the gravitational potential $\Phi = \Phi_{\mathrm{g}}$ and thus every particle does not feel the gravity of other particles while still feeling the gravitational acceleration from the gas potential. For the simulation indicated as 'no particle self-gravity' (simluation suffix \texttt{nopsg}), the dust particles neither contribute to the potential nor feel the gas potential and thus the particles only feel the aerodynamic drag force, the shearing box inertial forces, and the gravitational force from the central star.

\section{Results}
\label{sec:results}

We summarize the simulation results in Table \ref{tab:turbresults} and describe the two main diagnostics for the radial viscous stress and vertical particle diffusion below.

\begin{figure*}
\centering$
\vcenter{\hbox{\includegraphics[width=0.48\textwidth]{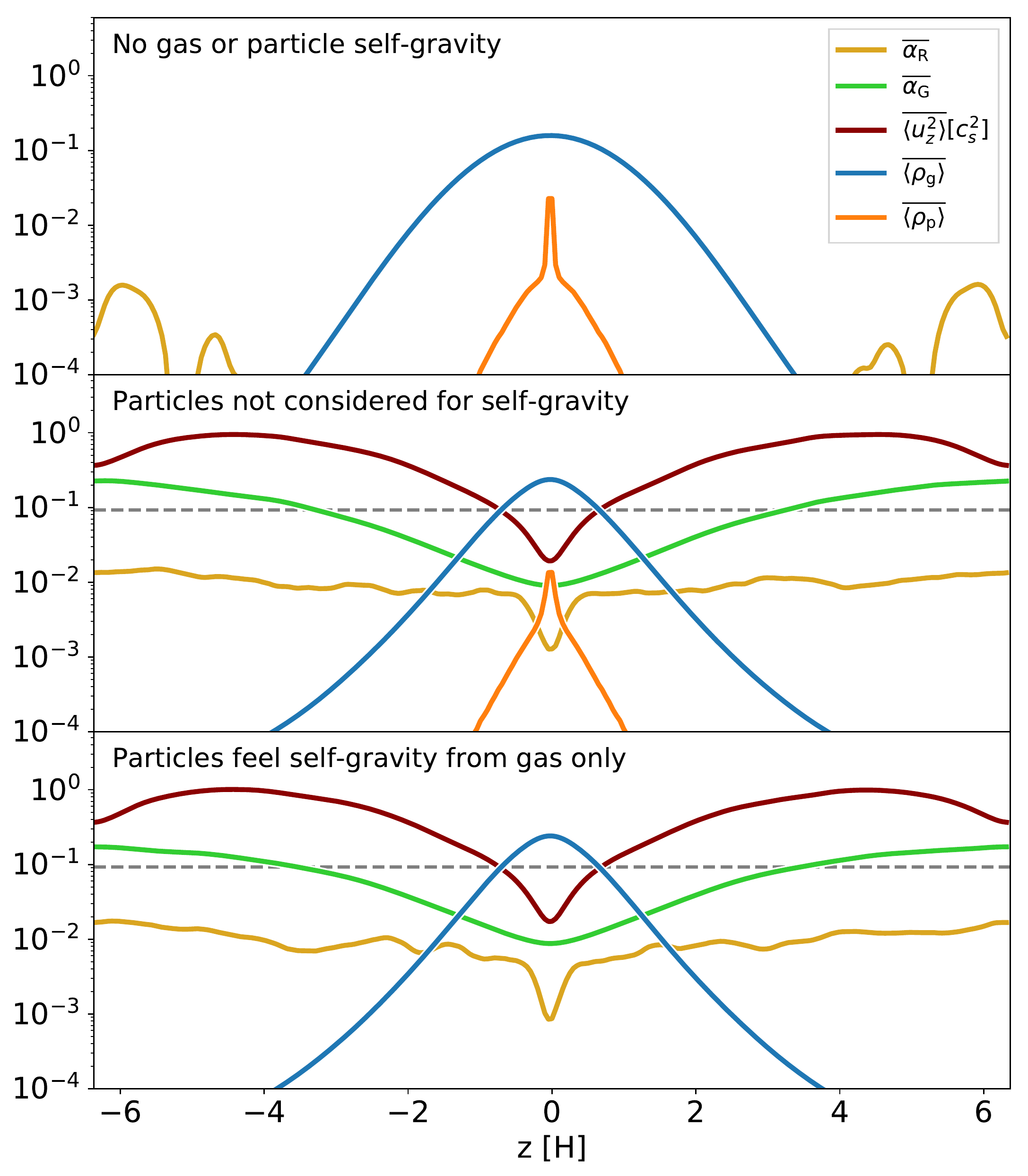}}}%
\vcenter{\hbox{\includegraphics[width=0.48\textwidth]{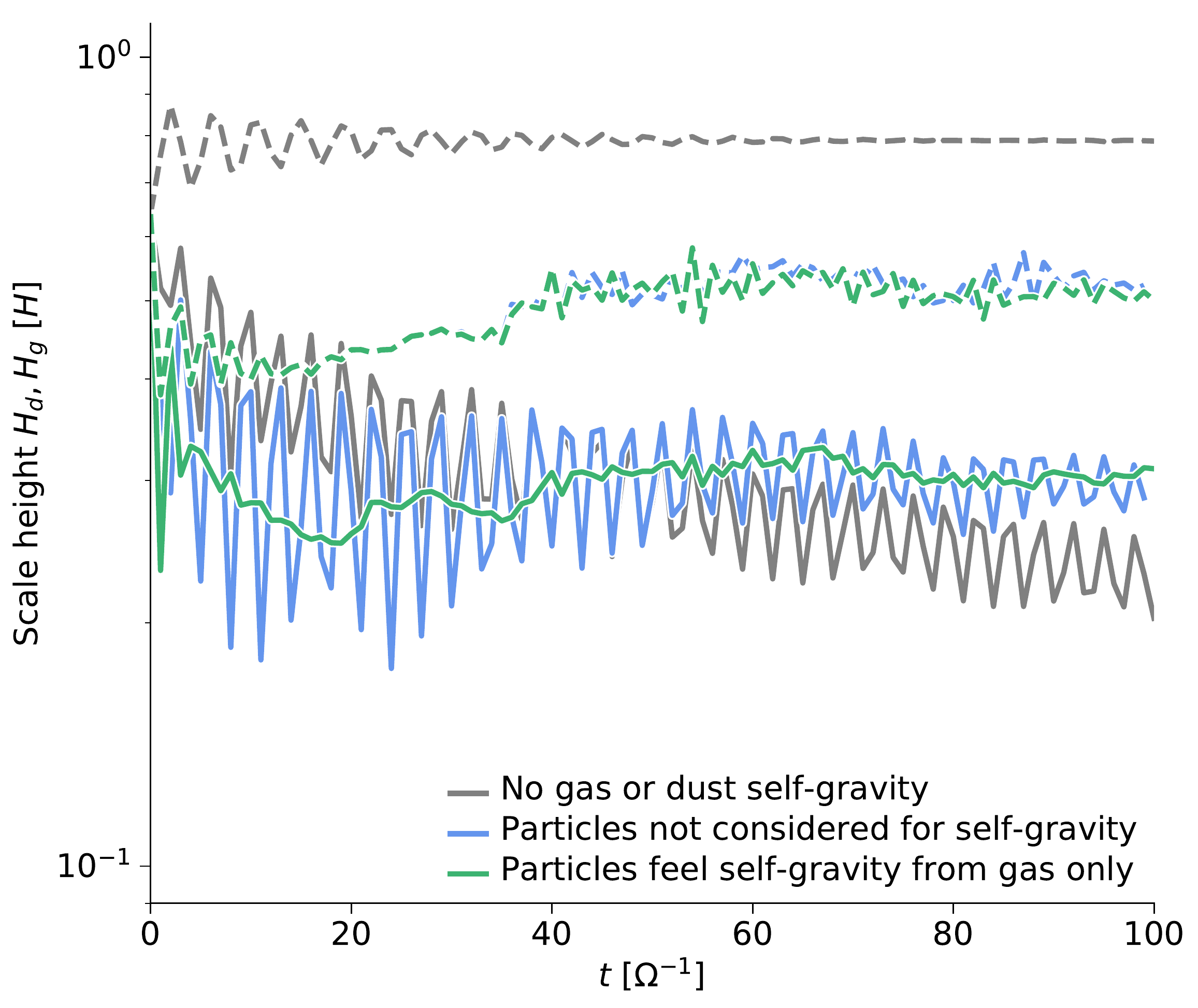}}}
$\caption{Comparison of three different implementations of particles. Left panel: vertical turbulence and density profiles as in Figure \ref{fig:densityturbulenceprofile}. Right panel: the evolution of the gas (dashed) and dust (solid) vertical scale heights. The oscillations in the particle scale height are due to the overdamped settling of the large dust species $\mathrm{St} \geq 10$. The first case does not include gas or dust self-gravity as a reference case, the second considers that particles do not contribute to the potential, i.e. $\Phi = \Phi_{\mathrm{g}}$, and do not feel the potential either. In the third case, the dust-to-gas ratio is smaller by a factor of $10^{-4}$, meaning they do not contribute significantly to the potential, but still feel the gas.}
\label{fig:selfgravitycomparison}
\end{figure*}

  \subsection{Turbulence}
  \label{subsec:turbulence}
  
We measure two properties of the gas turbulence: the $\alpha$ stresses and the dispersion of the vertical gas velocities $u_{z}^{2}$, averaged in the $x$-$y$ plane at different heights. They are presented in Table \ref{tab:turbresults} and Figure \ref{fig:densityturbulenceprofile}. Figure  \ref{fig:densityturbulenceprofile} also shows the gas and dust vertical density profiles in blue and orange lines, respectively. Turbulence may have transient fluctuations which produce negative stresses, even while averaged in the $x$-$y$ plane. To eliminate these fluctuations, quantities are averaged in time for 30 snapshots between $t=50\,\Omega^{-1}$ and $t=80\,\Omega^{-1}$, when turbulence has settled, and thus they are denoted with bars, e.g. $\overline{\langle u_{z}^{2} \rangle}$ or $\overline{\alpha}$. Even so, Reynolds stresses at the midplane are still lower than elsewhere and even negative as in the middle-right panel of Figure \ref{fig:densityturbulenceprofile}.

We compare the total stress of Equation \eqref{eq:totalalphastress}, as a simple volume average which can be compared with the analytic expression from \citet{Gammie2001}:
\begin{equation} \label{eq:gammiealpha}
\alpha = \frac{4}{9}\frac{1}{\gamma(\gamma - 1)\beta}.
\end{equation}
In Figure \ref{fig:densityturbulenceprofile}, Equation \eqref{eq:gammiealpha} is indicated by the gray dashed line. Averaged over the vertical extent, the gravitational stress (green line) is roughly an order of magnitude greater than the Reynolds stress. From the top to bottom panels, we can see that the ratio between $\alpha_G$ and $\alpha_R$ decreases ($\alpha_G/\alpha_R$ from 30 to 5) with a longer cooling time ($\beta$ from 2 to 10).  Thus, gravitational stress is more important for a faster cooling GI disk. 

The dark red lines in Figure \ref{fig:densityturbulenceprofile} show the vertical profiles of the vertical gas velocities and are typically subsonic, but can reach transonic values in the upper disk atmospheres. The averaged gas velocities in Table \ref{tab:turbresults} are calculated both as a simple volume average and a density-weighted volume average. The two differ by around $\sim 15\%$, since the dense midplane generally has lower vertical velocities, and the density-weighted average will lean towards the lower velocities near the midplane.

The particle density distributions all show some settling as a narrower peak near the midplane. The following section breaks down this particle distribution by size to determine how settling varies with each particle species.

  \subsection{Vertical Particle Diffusion}
  \label{subsec:verticalparticlediffusion}
  
In Figure \ref{fig:agglomMultispecies}, the vertical distribution of the four smallest dust species are shown. Particles are binned according to their nearest grid cell in Figure \ref{fig:densityturbulenceprofile}, and then summed over 30 snapshots. Particles with smaller Stokes numbers are more broadly distributed compared to larger particles, and will take tens of orbital periods to fully settle. Larger species $\mathrm{St} \geq 10$ settle as overdamped oscillations, overshooting the midplane many times in each direction before eventually settling. The especially large dust species with $\mathrm{St} \geq 100$ will do so for longer than the duration of the simulation and therefore are not included in the following analysis. Dust settling is balanced by the vertical diffusion from gravitoturbulence. Particles with $\mathrm{St}$ closer to 1 have faster settling which is harder for turbulent diffusion to balance, thus forming a narrower distribution.

\begin{figure*}
\centering
\includegraphics[width=0.48\textwidth]{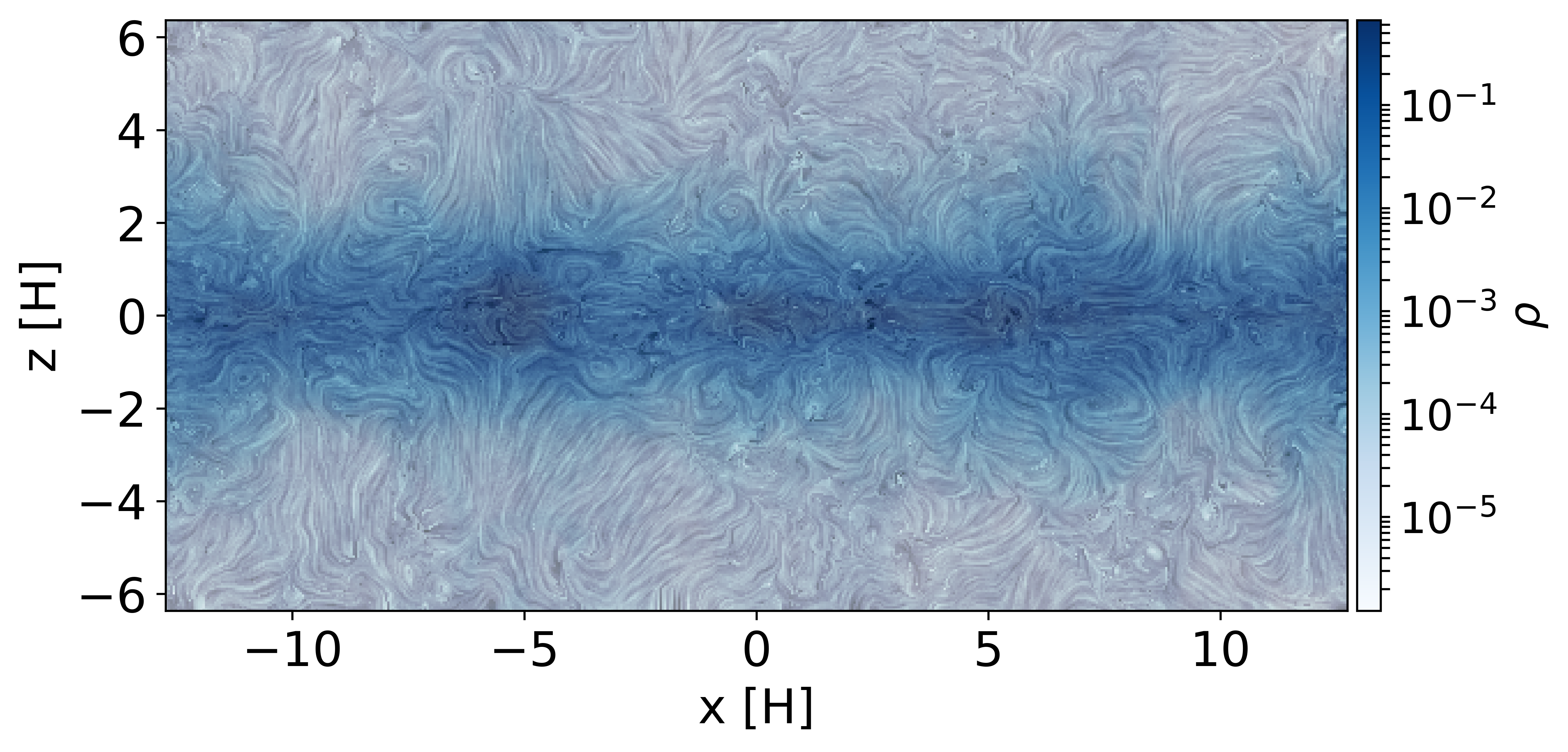}%
\includegraphics[width=0.48\textwidth]{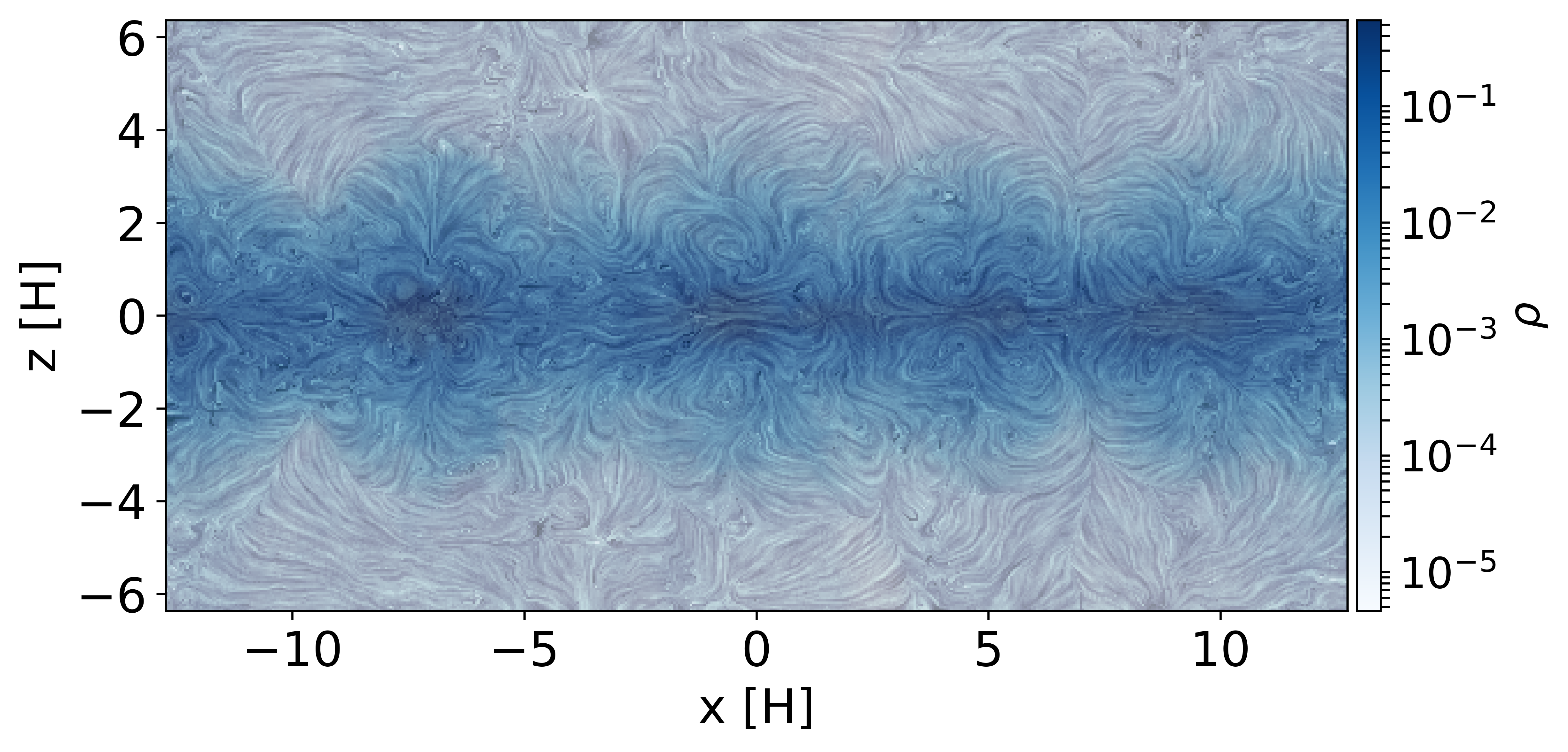}
\includegraphics[width=0.48\textwidth]{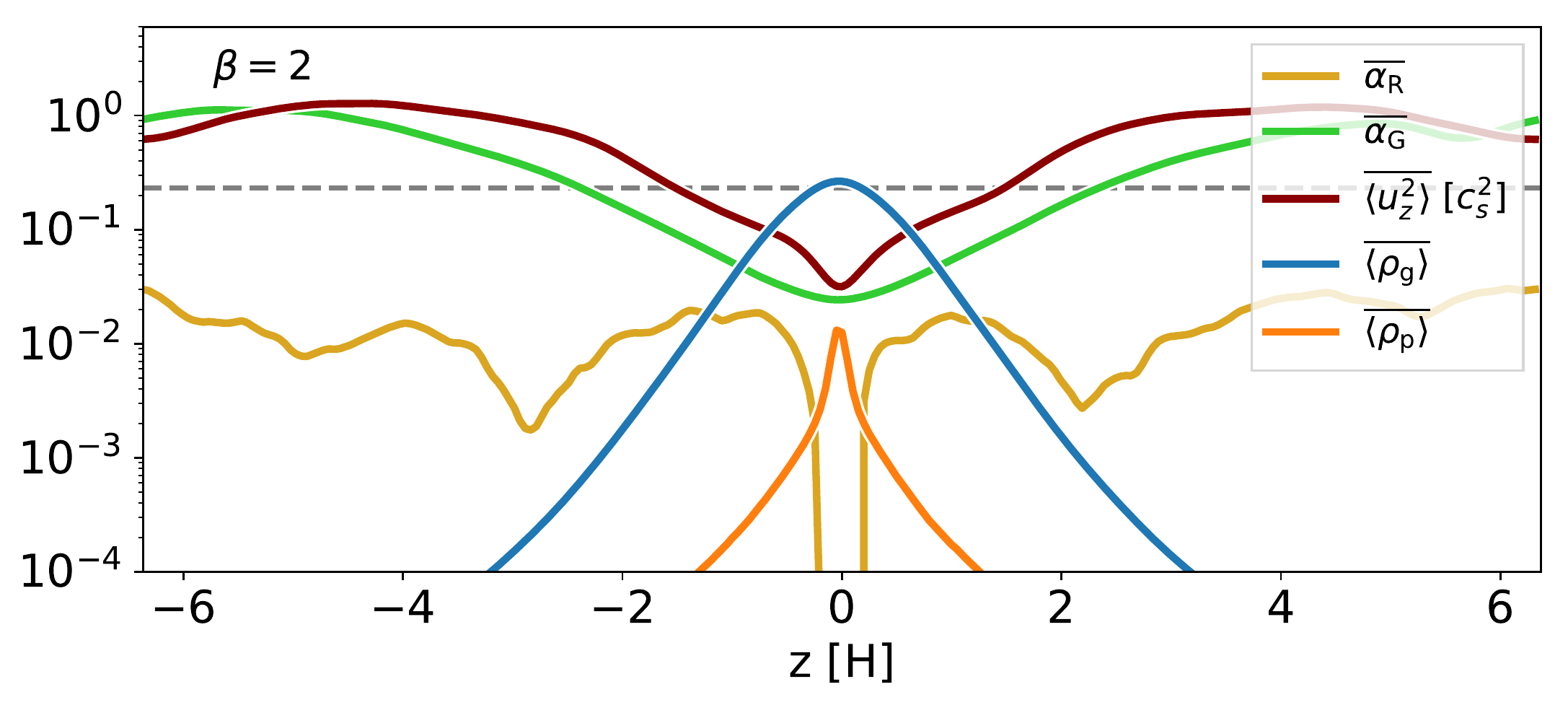}%
\includegraphics[width=0.48\textwidth]{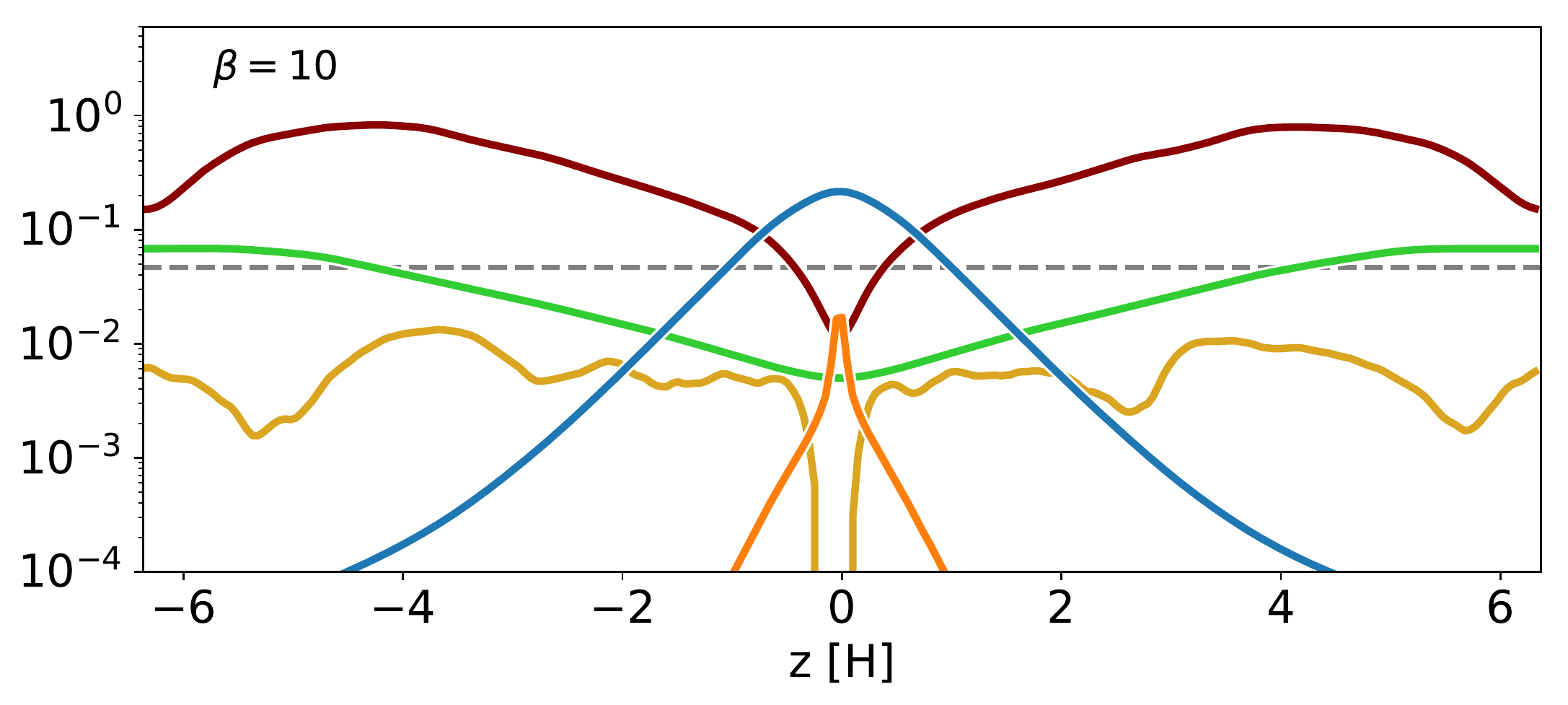}
\caption{Comparison of the $\beta=2$ and $\beta=10$ simulations near the end of the computation. Top two figures show a slice of the radial-vertical gas flow field using line integral convolution \citep{Cabral1993} at $t = 80\,\Omega^{-1}$. Bottom two figures show the averaged vertical profiles for only the last 5 snapshots before $t = 80\,\Omega^{-1}$.}
\label{fig:parametricinstabilitycomparison}
\end{figure*}

Based on the distribution of particles at each size, we determine the scale height for each dust species. We fit Gaussian profiles to the entire vertical distribution (i.e. $z = [-6H,6H]$) of each species binned to the nearest grid cell, using an initial guess with a width of $z=H$. In Figure \ref{fig:diffusionparameter}, we plot these particle scale heights against the inverted particle size $\mathrm{St}^{-1}$ in order to estimate the vertical particle diffusion coefficient $D_{d,z}$,
\begin{equation}\label{eq:diffusionparameter}
H_{\mathrm{d}} = \sqrt{\frac{D_{\mathrm{d,z}}}{\Omega\, \mathrm{St}}}\,.
\end{equation}
Equation \eqref{eq:diffusionparameter} is most appropriate for particles that have significant settling. Thus, we only use the scale heights of $\mathrm{St}=0.1$ and 1 in Figure \ref{fig:diffusionparameter} to derive the diffusion coefficients. The dimensionless diffusion constant $D^{'}_{\mathrm{d,z}} = D_{\mathrm{d,z}}/c_{\mathrm{s}} H$ is reported in Table \ref{tab:turbresults}. Alternatively, we can use a more general equation  which applies to both small and big dust to estimate the diffusion coefficient with all the three smallest species. By doing so, we find very similar values (see Appendix Section \ref{sec:appendixdiffusionconstant}). From the diffusion constant and $\alpha$ stress we define the vertical Schmidt number \citep{Dullemond2004}, a measure of the relative strength of the radial angular momentum transport compared to the vertical dust diffusion,
\begin{equation} \label{eq:schmidtnumber}
\mathrm{Sc}_{z} = \frac{\alpha}{D^{'}_{d,z}}\,.
\end{equation}
Traditionally, values of $\mathrm{Sc}=1$ are indicative of isotropic turbulence while values greater than unity may suggest that the radial turbulent motion (more strictly speaking, the $r\phi$ component of the turbulent stress) is stronger than the vertical turbulent motion. However, in GI disks, the turbulent stress $\alpha_R$ can be much weaker than the gravitational stress $\alpha_G$ due to disk self-gravity. Thus, we define $\mathrm{Sc}_z$ both in terms of the total stress as above and in terms of the Reynolds stress $\mathrm{Sc}_{R,z} = \alpha_{R} /D^{'}_{d,z}$.

Table \ref{tab:turbresults} suggests that $\mathrm{Sc}_{R,z}$ of the large simulations calculated with the Reynolds stresses are close to unity, while $\mathrm{Sc}_{R,z}$ from the medium-sized simulations are around a factor of 4 greater. Due to the large ratio between $\alpha$ and $\alpha_R$ in our GI simulations, the Schmidt number calculated with the total $\alpha$ stress is much larger, almost two orders of magnitude greater in some cases. This indicates a significant amount of settling of the intermediate-sized dust compared to the high radial accretion.

We highlight an observed trend in the ratios $\alpha/\alpha_{R}$ and $\mathrm{Sc}_{z}$ as function of the cooling timescale in Figure \ref{fig:turbulentratios}. With rapid cooling and strong radial transport, both $\alpha/\alpha_{R}$ and $\mathrm{Sc}_{z}$ are very high, but decrease as the gravitoturbulence weakens and becomes more comparable in strength to the Reynolds stress. We expect that simulations with longer cooling times, i.e. $\beta \sim$ 20-30, and resolution similar to or medium boxes could have Schmidt number that explain the turbulence levels and dust settling of disk observations.

In Figure \ref{fig:scaleheightevolution} we show the evolution of the dust and gas scale heights over the duration of the simulation. The gas pressure scale height, indicated by the dashed lines, is less than $H_{g}/H=1$ for two reasons. First, since we use an adiabatic equation of state, the gas scale height is $H_{g} = H/\sqrt{\gamma} = 0.77 H$. Second, gas self-gravity has a small, but noticeable effect on the gas distribution, such that it is no longer Gaussian and slightly narrower within two scale heights of the midplane \citep{Riols2017}. This discrepancy can be seen in the right panel of Figure \ref{fig:selfgravitycomparison}, where the gas in a disk without self-gravity has settled to $H_{g}/H \sim 0.8$ and the others with self-gravity are narrower, as indicated by the blue lines in the panels to the right.

\begin{figure*}[t]
\centering$
\vcenter{\hbox{\includegraphics[width=0.48\textwidth]{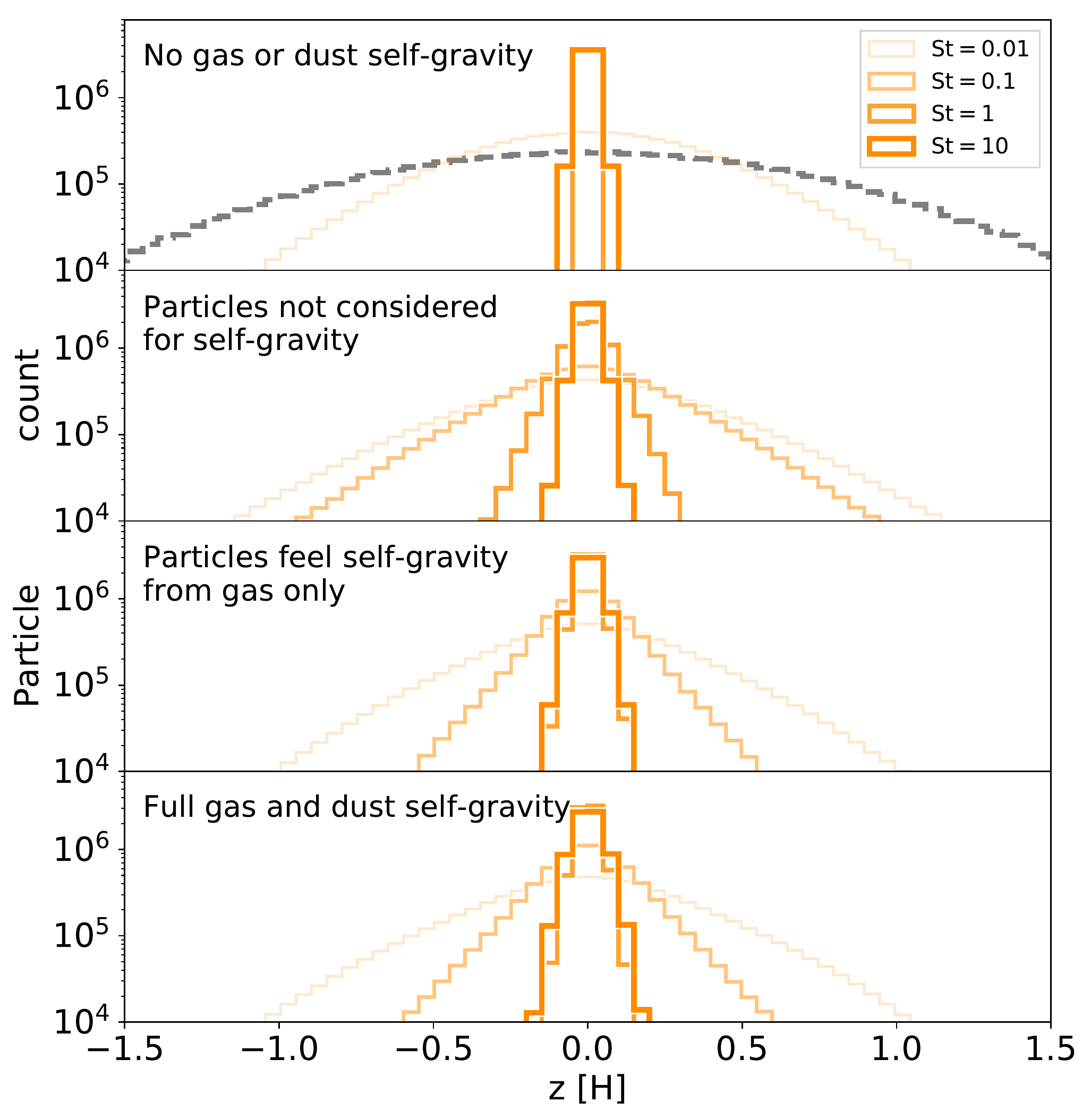}}}%
\vcenter{\hbox{\includegraphics[width=0.48\textwidth]{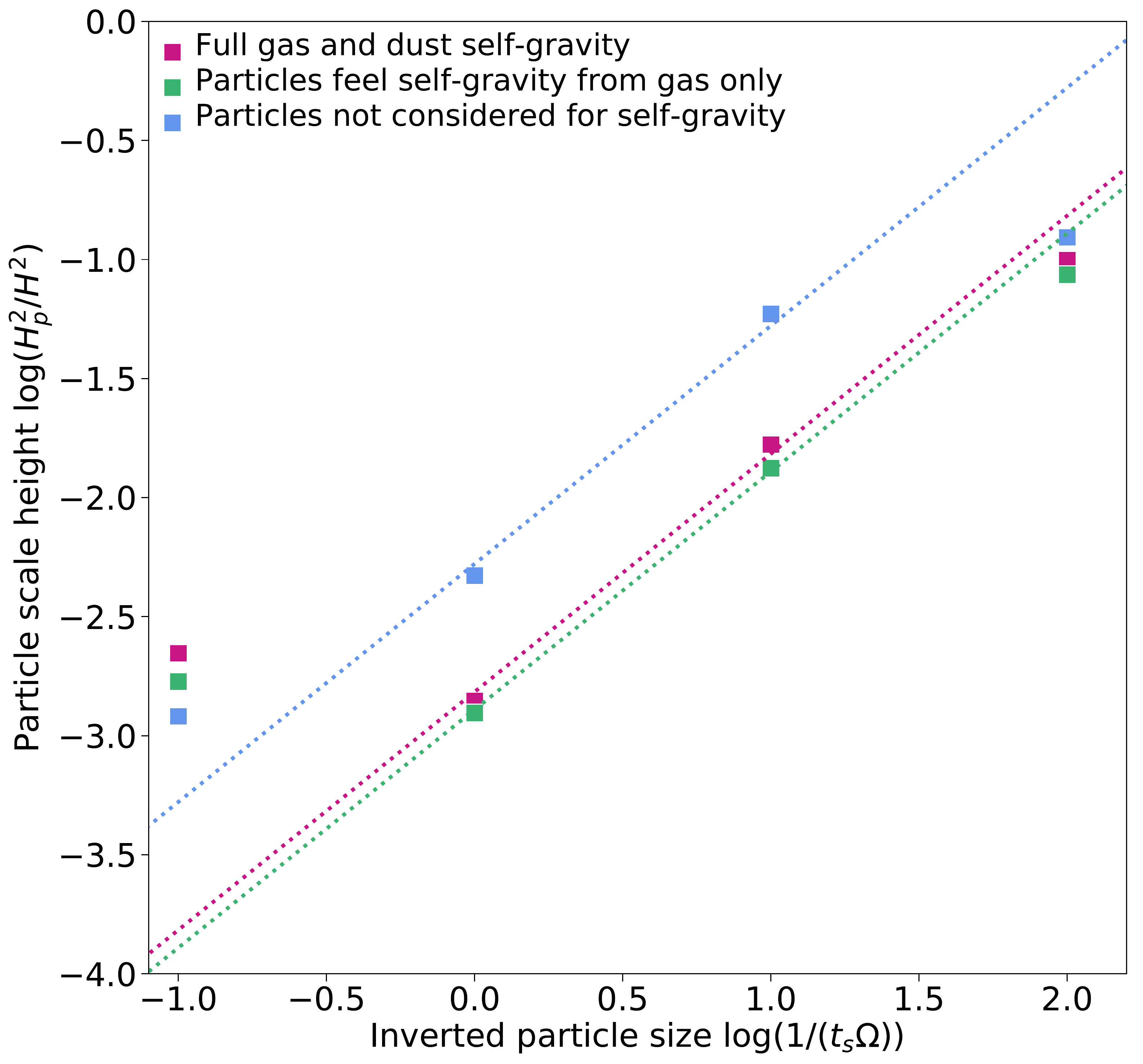}}}
$\caption{Left panel: For the additional simulations which explore the role of self-gravity on particle settling, the vertical distribution of each of the four particle species as in Figure \ref{fig:agglomMultispecies}. Right panel: the particle scale height versus particle size, from which the vertical diffusion constant of the dust $D_{d,z}$ is derived using Equation \eqref{eq:diffusionparameter}.}
\label{fig:selfgravitycomparison2}
\end{figure*}

\begin{figure}[t]
\centering
\includegraphics[width=0.48\textwidth]{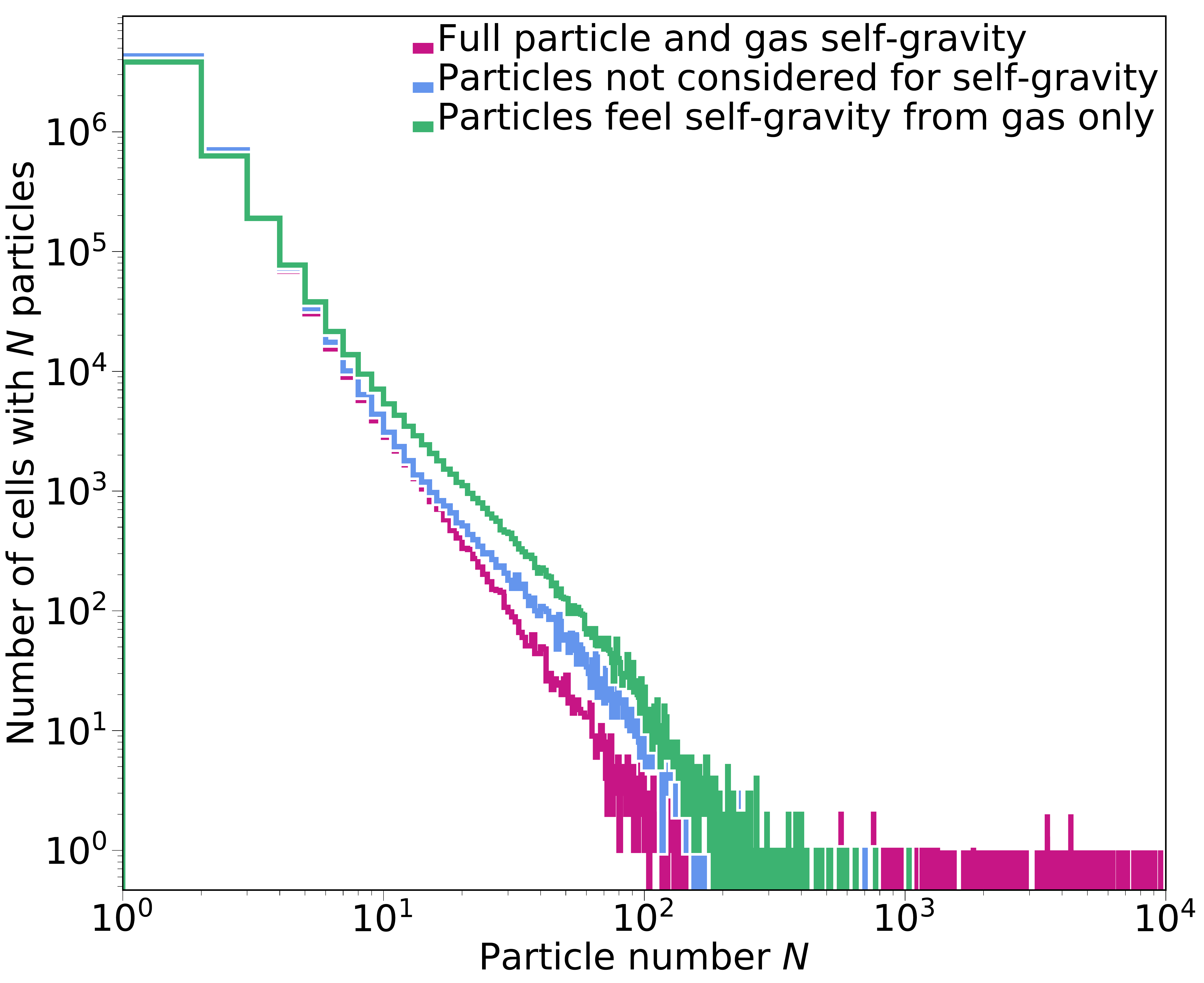}
\caption{Histogram of the number of cells with $N$ particles in the three simulations which evaluate particle self-gravity. A lower dust-to-gas ratio $\epsilon$ keeps dense filaments of dust ($N\sim 10^{2}$) from collapsing into clumps ($N\gtrsim 10^{3}$). When particles feel no effects from self-gravity, concentrations are greatly diminished.}
\label{fig:concentrations}
\end{figure}

The dust scale height including all particle species is indicated by the solid line in Figure \ref{fig:scaleheightevolution}. In both medium-sized and big box simulations, the dust scale heights with the most efficient cooling $\beta=2$, increase consistently over a long time scale until it matches the gas scale height. This appears to correspond to some excess Reynolds stress just above the the disk midplane (see gold lines in top two panels of Figure \ref{fig:densityturbulenceprofile}), potentially an indicator of the parametric instabilities reported in \cite{Riols2017}.

A parametric instability in a self-gravitating disk would be the result of an axisymmetric mode interacting with inertial waves with half the frequency. We investigate this further in Figure \ref{fig:parametricinstabilitycomparison}. Apparent in Figure \ref{fig:scaleheightevolution} is the gradual increase of the particle scale height with time in simulations as long as $\beta=2$, perhaps indicating the presence of additional turbulence. Figure \ref{fig:parametricinstabilitycomparison} shows a comparison of the gas flow patterns in the top row and the vertical properties closer to the end of the end of the simulations. When $\beta = 10$ the symmetric circulation patterns near the intersection with the high-density spiral arms is stronger compared to the case where $\beta=2$ and the pattern is disrupted by turbulence near the midplane (compare with Figure 9 of \citet{Riols2018a}). In the vertical profile of the latter, the vertical motions around the midplane $u_{z}^{2}$ are suppressed.

\section{Discussion}
\label{sec:discussion}

The properties of turbulence in our simulations are comparable with those of \citet{Shi2014}, which also focus on 3D gravitoturbulence, but do not include particles. Similarly, we find that a simple volume average of the $\alpha$ most closely matches the analytic results of Equation \ref{eq:gammiealpha}. The vertical gas velocities $u^{2}_{z}$ are stronger at a few scale heights above the disk midplane from the subsonic to transonic regions, as with \citet{Riols2017}.

We emphasize the importance of the gas self-gravity acting upon particles in Figure \ref{fig:selfgravitycomparison2}. The left panel shows the vertical distribution of particles calculated based on $30$ snapshots from $t=50\,\Omega^{-1}$ to $t=80\,\Omega^{-1}$, but with different treatments for self-gravity. From these simulations we draw a number of key conclusions about how self-gravity affects the settling of dust and what it implies about the vertical diffusion.

\subsection{Particle Settling}
\label{subsec:particlesettling}

\begin{figure*}
\centering
\includegraphics[width=0.48\textwidth]{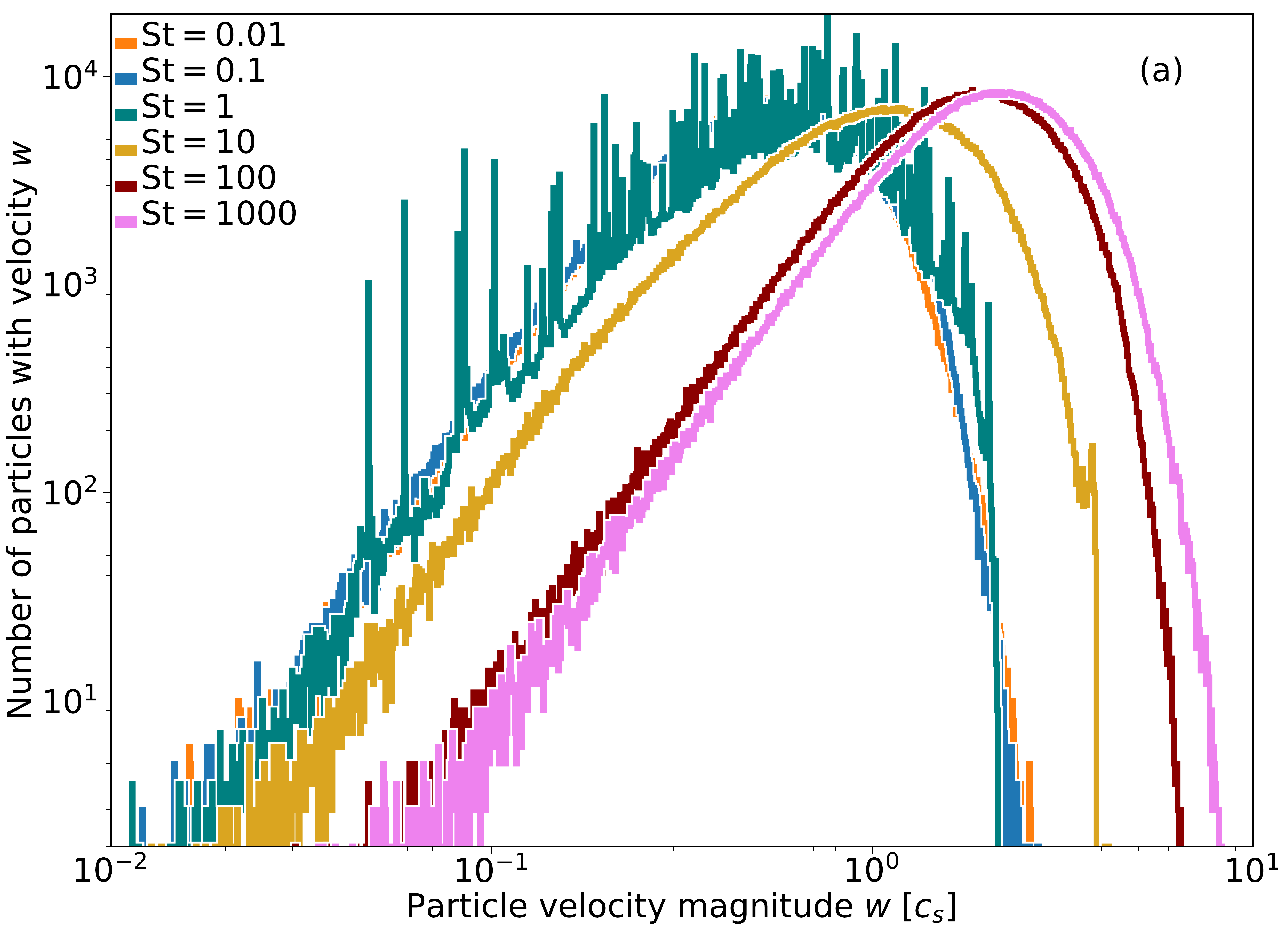}%
\includegraphics[width=0.48\textwidth]{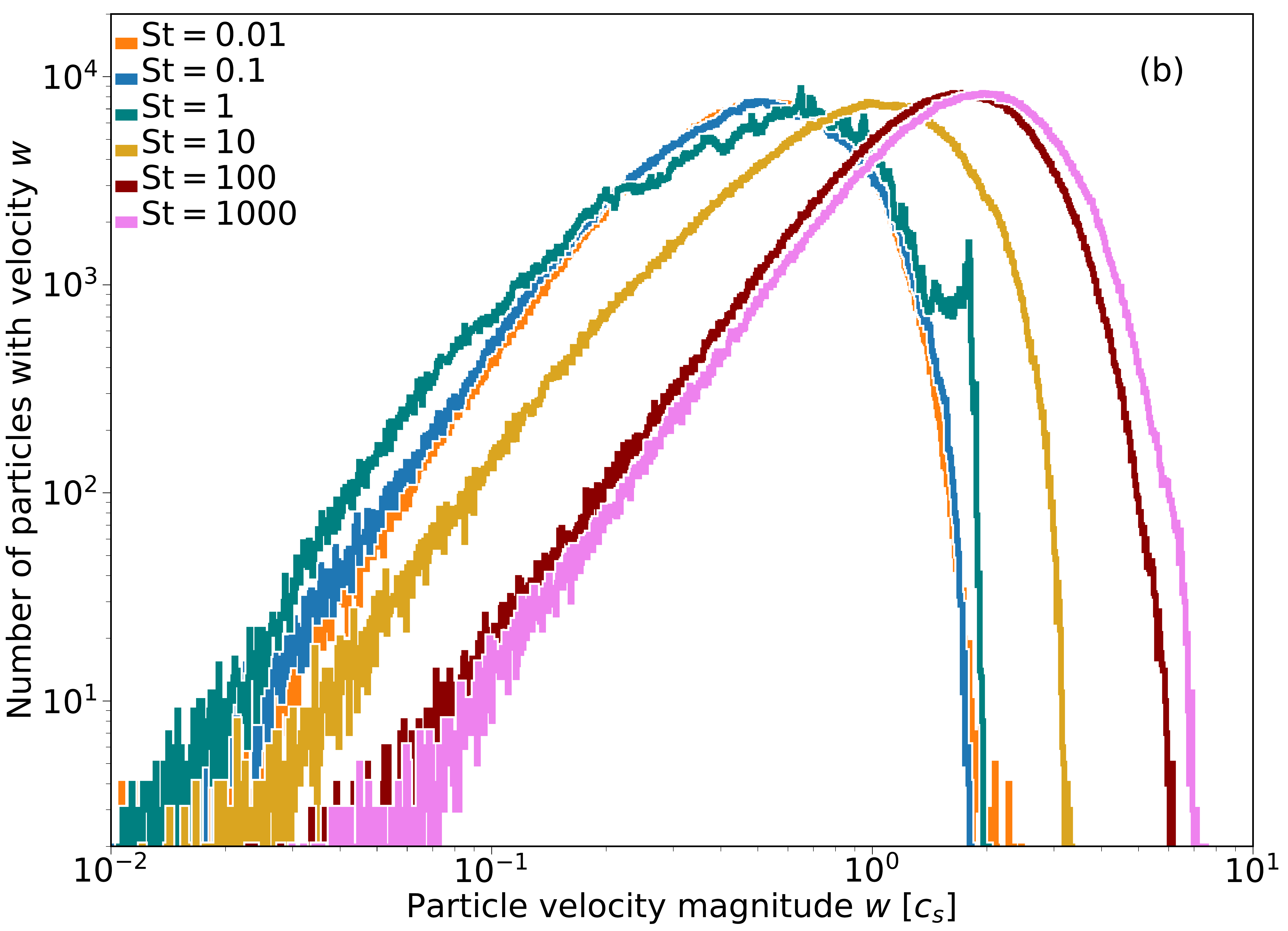}
\caption{Distribution of particle velocity magnitudes in terms of the gas sound speed $c_{s}$ for the two cases where particle self-gravity is included, but with (a) more massive particles ($\epsilon = 10^{-2}$) and (b) less massive particles ($\epsilon = 10^{-6}$). When the self-gravity between particles is non-negligible, particles $\mathrm{St=1}$ can concentrate into clumps of up to a thousand with low relative velocities, indicated by the spikey features in (a).}
\label{fig:velocityhistogram}
\end{figure*}

The top panel in Figure \ref{fig:selfgravitycomparison2} shows a settling test of the particles in the complete absence of any self-gravity from either the gas or the particles. Thus, particles settle in a static disk without turbulence. The settling time is inversely proportional to their sizes $t_{\mathrm{settle}} \sim 1/\mathrm{St}$ for $\mathrm{St}\leq 1$ particles \citep{Youdin2007a} in the absence of any turbulent diffusion. The intermediate particles $\mathrm{St} = 0.1,1,10$ settle rapidly and form a thin layer at the two grid cells on either side of $z=0$. The simulation does not last long enough for $\mathrm{St} = 0.01$ particles to completely settle.

The second panel from the top in Figure \ref{fig:selfgravitycomparison2} shows how particles settle in a gravitoturbulent disk with only the gas component being considered for self-gravity. Particles only feel the vertical gravity of the stellar potential and do not feel the gravity from the gas or dust mass concentrated at the disk midplane. Thus, without any self-gravity acting on the dust, each particle size has a thicker layer compared to the cases which include self-gravity. This test studies the dust diffusion by the gravitoturbulence solely without including additional concentration effects due to the gas and dust gravity. The two intermediate sized particle species ($\mathrm{St}=$0.1, 1) are diffused by the gravitoturbulence of the gas while the largest and smallest particles are largely unchanged. 

In the third panel from the top, the same simulation is run again, but particles do feel the gravitational acceleration of the gas, but not of the other particles. Here we see noticeable settling of the intermediate dust and slight narrowing of the distribution for the smallest dust. The narrowing is roughly consistent with our expectation (Equation \ref{eq:selfgravitydust}). Finally in the bottom panel, the dust particles contribute to the self-gravity potential and is fully self-consistently included, but it has little effect on the overall settling of any dust species compared with the panel above.

The dust scale heights and the diffusion coefficients for these tests  are shown in the right panel of Figure \ref{fig:selfgravitycomparison2} and Table \ref{tab:turbresults}. Comparing the $\texttt{S\_t5\_B\_nopsg}$ and $\texttt{S\_t5\_B\_lowpsg}$ cases, we see that the gas gravity reduces the dust scale height by a factor of $\sim 1.8$. The turbulence is quite isotropic ($\mathrm{Sc}_{R,z}=1.7$) when the gas gravity is neglected (the $\texttt{S\_t5\_B\_lowpsg}$ case). Strictly speaking, when we include the gravity from gas to dust, the turbulent properties are not changed so that we should use the same diffusion coefficient and Equation \ref{eq:selfgravitydust} to derive the dust scale height. On the other hand, it is more convenient to factor the $Q$ coefficient of Equation \ref{eq:selfgravitydust} into the definition of dust diffusion coefficient and use Equation \ref{eq:diffusionparameter} to derive a new dust diffusion coefficient. This is also how previous observations constrain the diffusion coefficient (e.g. \citealt{Pinte2016}). Thus, the new dust diffusion coefficient defined in this way is reduced by a factor of $\sim$3, so that $\mathrm{Sc}_z$ and $\mathrm{Sc}_{R,z}$ is increased by a factor of $\sim$3 compared with the case without gas gravity.

\subsection{Vertical Schmidt Number}
\label{subsec:verticalschmidtnumber}

Such a high Schmidt number (20-200) is crucial for explaining the moderate disk accretion \citep{Ribas2020} but thin dust disk \citep{Villenave2020}. \cite{Pinte2016} estimate that $Sc_z\ge 10$ is needed to explain HL Tau's accretion rate and thin dust disk. 

The Schmidt number has been studied for turbulence driven by several instabilities in protoplanetary disks, including the magneto-rotational instability (MRI) \citep{Johansen2005a,Carballido2011,Zhu2015,Riols2018} and the vertical shear instability (VSI) \citep{Stoll2016}. Simulations of the MRI show that the particle diffusion is comparable to the angular momentum transport, i.e. Schmidt numbers around unity. In some cases, especially for non-ideal MHD, the Schmidt number can be even smaller than 1 \citep{Zhu2015,Yang2018}. In the case of VSI, \citet{Flock2020} find that Schmidt number of VSI is $\sim0.04$ due to the thin vertical circulating motion. Thus, both MRI and VSI cannot explain observations. Other accretion mechanisms need to be explored, such as disk wind \citep{Riols2018} or disk self-gravity.

Some previous works have explored how dust in self-gravitating disks can settle. \cite{Sengupta2019} use a Monte Carlo turbulent settling model and find that dust settling is enough to affect the local Toomre stability of the disk. But it does not include the gravitational force from the gas disk to the dust. \citet{Riols2020} use much the same methods as here, but arrive at the conclusion that settling is not very significant. This may in part be due to the fluid description of solids which limits the range of included dust particles to $\mathrm{St}<1$. For $\mathrm{St} = 0.1$ at $\beta=20$ we find the ratio of the dust-to-gas scale height $\overline{H_{d}/H_{g}} \sim 0.23$, similar to the value for $\mathrm{St} = 0.16$ in Figure 3 of \citet{Riols2020}. However, whereas their value for vertical diffusion $D_{z} = 0.013$ predicts the scale height ratio to be $H_{d}/H_{g}\sim 10^{-1}$ at $\mathrm{St} = 1$, we find scale height ratios as low as $\overline{H_{d}/H_{g}}=5 \times 10^{-2}$. Thus, we infer a lower diffusion constant. This trend continues in our simulations at $\beta=20$ and $\beta=30$ and our Schmidt numbers remain high, suggesting more research is needed to quantify the strength of dust settling. Nevertheless, based on the diffusion and accretion coefficients provided in \citet{Riols2020}, $Sc_{z}$ is $\sim$4 in their simulations, which is also larger than 1, in broad agreement that non-isotropic diffusion is operating in GI disks. 

\begin{figure}
\centering
\includegraphics[width=0.48\textwidth]{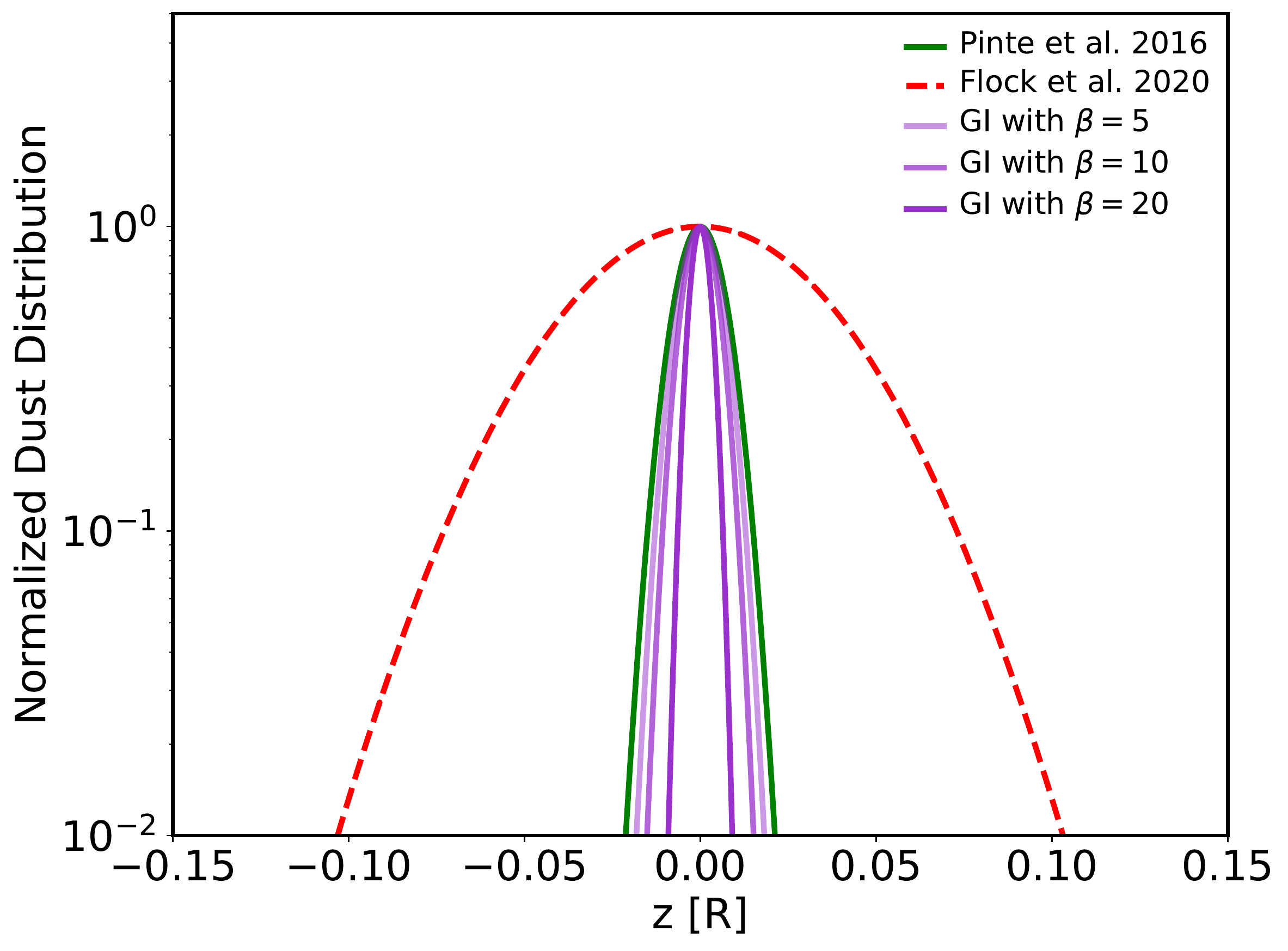}
\caption{Comparison of the VSI settling in \citep{Flock2020} (see their Figure 9) with the settling expected using Equation \eqref{eq:diffusionparameter} assuming $\mathrm{St}=5\times 10^{-2}$ and using the diffusion values in Table \ref{tab:turbresults} for the medium-sized boxes at $\beta=5,10,20$.}
\label{fig:Flock2020}
\end{figure}

Our simulations suggest that gravitationally unstable disk can easily have $Sc_z\ge$10, consistent with HL Tau observations. Furthermore, due to the early evolutionary stage of HL Tau, the disk is quite massive \citep{Carrasco-Gonzalez2019} and GI will naturally operate in such a disk. In Figure \ref{fig:Flock2020}, we compare the dust settling which results from some of the diffusion parameters measured in Table \ref{tab:turbresults} with the settling inferred from HL Tau observations \citep{Pinte2016} and from global VSI simulations \citep{Flock2020}. All the GI cases shown have accretion $\alpha>0.01$. Observations indeed suggest significant accretion equivalent to $\alpha\approx 0.01$ at 100 au \citep{Beck2010,Pinte2016} which is not congruent with the angular momentum transport typically associated with VSI, but may be consistent with MRI \citep{Flock2017}. Gravitoturbulence on the other hand, can provide both significant accretion along with a thin dust disk. Unexplained however is why the spirals characteristic of such a massive disk would be absent in ALMA dust continuum observations. This can be due to some other mechanisms which are producing gaps in HL Tau and concentrate dust in rings. In this case, the spirals may still be present in the gas. Or, the lack of observed spirals can be due to that the $Q$ in the disk is not low enough to trigger the spirals.

We emphasize that even if $Q>1$ and the disk is gravitationally stable, the additional settling due to gas gravity cannot be neglected (Equation \ref{eq:selfgravitydust}). At a height $z=H$ above the disk midplane, vertical acceleration due to stellar gravity is comparable to self-gravity when
\begin{equation} \label{eq:gravitycomparison}
\Omega^{2}H = 2\pi G\Sigma
\end{equation}
which corresponds to $Q=2$. Thus, the effects of disk self-gravity are likely to be non-negligible for $Q$ up to a few. Future simulations will need to focus on the interaction between gas self-gravity and VSI, as even a $Q=2$ disk may be enough to reduce the dust scale height in a VSI disk significantly. HL Tau has a measured Toomre $Q$ as low as 2 \citep{Liu2017}, which should be not expected to produce spiral features as long as $Q\gtrsim 1.7$, consistent with the ring features of HL Tau. Overall, disk self-gravity may still promote dust settling even in a gravitationally stable disk that is turbulent by some other mechanisms, including VSI.

\subsection{Particle Self-Gravity}
\label{subsec:particleselfgravity}

Gravitoturbulent disks have been suggested as a possible location for planetesimal formation \citep{Rice2004,Gibbons2012} as long as relative velocities between dust particles are low enough to avoid fragmentation events which prevent growth \citep{Booth2016}. While the gravitational acceleration between particles (dust disk's self-gravity) has little effect on vertical dust settling, it is only with particle-particle self-gravity that dense clouds of dust are formed. In Figure \ref{fig:concentrations}, we show the number of grid cells with $N$ particles for three comparable simulations, aside from the implementation of self-gravity. Only in the case where particles can feel the self-gravity of the other particles are there a considerable number of concentrations up to $N=10^{4}$. When the dust does not feel any self-gravity at all (blue line), particles concentrate less in the intermediate ranges $N \sim 10 - 100$ compared to the simulation with gas self-gravity acting on the particles (green line).

We plot the velocity distribution of each particle species in Figure \ref{fig:velocityhistogram} for the case where (a) particle-particle self gravity is included and (b) when dust mass is too low for it to matter. The sharp features of the $\mathrm{St}=1$ in (a) indicate that many of the dense clumps from Figure \ref{fig:concentrations} are groups of intermediate-sized particles with low relative velocities and potentially bound. Whether these particle clouds can become planetesimals depends on whether or not dust can grow to the intermediate sizes fast enough and avoid encountering the fragmentation and drift barriers of planetesitmals \citep{Booth2016}. The critical length scale to be resolved for collapsing to a planetesimal is \citep{Klahr2020}
\begin{equation} \label{eq:collapselengthscale}
l_{c} = \frac{1}{3}\sqrt{\frac{D^{'}}{\mathrm{St}}}H\,,
\end{equation}
which, for the diffusion constants we measure $D^{'} \sim 10^{-3}$, is too small ($l_{c} \sim 10^{-2}\, H$) to be resolved in our simulations. Even if a simulation is unable to resolve free-fall collapse to planetesimals, dense dust clouds at or above Hill density
\begin{equation} \label{eq:hilldensity}
\rho_{\mathrm{Hill}} = \frac{9}{4\pi}\frac{M_{*}}{R^{3}}\,,
\end{equation}
can be subject to diffusion-regulated collapse \citep{Gerbig2020}, given a stellar mass $M_{*}$ and an orbital radius $R$. Several of our $\mathrm{St}=1$ dust clouds are above dust-to-gas ratio of unity and at Hill density. Whether these clouds have the sufficient conditions for further collapse in the presence of the large scale diffusion of gravitoturbulence and the smaller scales turbulence due to streaming instability will be studied in future work.

We note that particle feedback, the effect of solid material pushing a fluid as it passes through it, is not included in these models. When local dust-to-gas ratios approach $\epsilon = 1$ and greater, feedback will become important and may lead to enhanced particle densities.

\subsection{Additional Considerations}
\label{subsec:additionalconsiderations}

We want to note that another explanation for well-settled dust disks comes from the launching of winds from the disk surface. A disk wind could remove angular momentum and provide moderate accretion without adding any turbulence to the midplane, allowing dust to settle \citep{Bai2013}. In tandem with a layered accretion, where disk surfaces are MRI active and the midplane is a dead zone due to non-ideal MHD effects \citep{Gammie1996a}, disk winds could work together to explain the high accretion and settling \citep{Hasegawa2017}. On the other hand, even if this is the case and the disk is gravitationally stable, the studied gravitationally effect to reduce dust scale height (e.g. Eq. \ref{eq:selfgravitydust}) should still operate in such massive disks.

While the box size does not appear to be a decisive factor in the results of our simulations, medium boxes are slightly more consistent with analytic predictions of overall turbulent stress $\alpha$. This is likely due to the better resolution, although the resolution of both domain sizes is within the converged ranges of similar studies. However, the gas density profile is slightly broader in these larger domains, and the lack of resolution in the vertical direction may have a two-fold effect. Slightly stronger Reynolds stresses would broaden the gas density profile and as a result weaken the vertical self-gravity. This can be seen in the almost order of magnitude difference between $\mathrm{Sc}_{R,z}$ of each size in Figure \ref{fig:turbulentratios}. With higher resolution, the Reynolds stresses appear to become less isotropic. \citet{Booth2019} find that gravitoturbulent diagnostics like $\alpha$ and the volume averaged $Q$ are unaffected by resolution down to 8 grid cells per scale height $H$ (see their Figure 1), for a simulation size slightly larger than the medium boxes used here. Our resolution is 10 grid cells per $H$ for the big boxes, and 20 cells per $H$ for the smaller boxes. These smaller boxes are similar in size $L_{x}=L_{y}=25H$ and effective resolution (20 grid cells per $H$) to the primary runs of \citet{Riols2020}. They see a lower resolution by a factor 2, which is comparable in resolution to our big box runs, does not change the measured dust scale heights significantly (their Figure 3). However, box size is still an important factor which determines which modes are incorporated in disk stability, reflected in the overall larger $Q$ values for larger boxes. \citep{Booth2019}. Ultimately, as an instability with long-range effects, we aim to model GI in global simulations, which will provide the most insight into dust settling in gravitoturbulent disks and compare with observations.

\section{Conclusion}
\label{sec:conclusion}

Using 3D vertically stratified local hydrodynamic simulations, we investigate the vertical settling of different sized dusts in gravitationally unstable disks. We systematically explore how different cooling times, different box sizes, and treatments of gas/dust gravity affect the particle settling. We summarize our results in the following points:

\begin{itemize}
    \item Turbulent strength in our simulations is consistent with the analytical expectation. The gravitational stress is comparable to the Reynolds stress near the midplane, but as the latter falls off with height, the volume averaged gravitational stress is stronger than the turbulent Reynolds stress by a factor of $5-30$, increasing with a shorter cooling time.
    \item Dust settling is enhanced by the vertical gravitational attraction of the gas onto the particles. Even when a disk becomes less massive and the disk is gravitationally stable (e.g. $Q>1$), the influence of gas self-gravity on particles should be included when studying dust settling in disks with other turbulent driving mechanisms.
    \item The resulting high Schmidt numbers due to both strong gravitational stress and gas vertical gravity ($\mathrm{Sc}_z \sim 20-200$) suggest gravitationally unstable disks could have strong accretion accompanied by thin dust disks. The high amount of settling could possibly help explain the observations of young, thin, dust protoplanetary disks accreting at moderate rates (e.g. HL Tau). 
    \item  While self-gravity of the gas onto the dust particles causes the significant settling of intermediate-sized dust, self-gravity between dust particles of the same size produces several dense clouds with high dust-to-gas ratios. These clouds have low relative velocities and could continue to collapse to form planetesimals at a very early stage in the disk lifetime.
\end{itemize}

Ultimately, self-gravity of the dense gas midplane has a significant affect on the vertical settling of dust in gravitoturbulent disks. Despite the strong turbulence and high radial diffusion typical of gravitoturbulent disks, the gravitational acceleration of the dense gas midplane causes significant settling of intermediate dust sizes. This could help to explain recent ALMA observations and also promote planetesimal formation.

\vspace{5pt}

\acknowledgments
The authors thank the anonymous referee for valuable feedback. HB thanks Henrik Latter, Chao-Chin Yang and Laith Mashni for valuable discussions. This research was supported by NASA TCAN award 80NSSC19K0639 and discussions with associated collaborators. Simulations made use of the Isaac cluster at the Max-Planck Center for Data and Computing in Garching and the Texas Advanced Computing Center (TACC) at the University of Texas at Austin through XSEDE grant TG-AST130002.

\software{Matplotlib \citep{Hunter2007}, SciPy \& NumPy \citep{Virtanen2020,vanderWalt2011}, IPython \citep{Perez2007}}

\bibliography{library}

\begin{thebibliography}{}
\expandafter\ifx\csname natexlab\endcsname\relax\def\natexlab#1{#1}\fi
\providecommand{\url}[1]{\href{#1}{#1}}
\providecommand{\dodoi}[1]{doi:~\href{http://doi.org/#1}{\nolinkurl{#1}}}
\providecommand{\doeprint}[1]{\href{http://ascl.net/#1}{\nolinkurl{http://ascl.net/#1}}}
\providecommand{\doarXiv}[1]{\href{https://arxiv.org/abs/#1}{\nolinkurl{https://arxiv.org/abs/#1}}}

\bibitem[{Baehr \& Zhu(2021)}]{Baehr2021}
Baehr, H., \& Zhu, Z. 2021, Particle Dynamics in 3D Self-gravitating Disks I:
  Spirals.
\newblock \doarXiv{2101.01888}

\bibitem[{Bai(2015)}]{Bai2015}
Bai, X.~N. 2015, The Astrophysical Journal, 798, 84,
  \dodoi{10.1088/0004-637X/798/2/84}

\bibitem[{Bai \& Stone(2013)}]{Bai2013}
Bai, X.-N., \& Stone, J.~M. 2013, The Astrophysical Journal, 769, 76,
  \dodoi{10.1088/0004-637X/769/1/76}

\bibitem[{Balbus \& Hawley(1991)}]{Balbus1991}
Balbus, S.~A., \& Hawley, J.~F. 1991, The Astrophysical Journal, 376, 214

\bibitem[{Beck {et~al.}(2010)Beck, Bary, \& McGregor}]{Beck2010}
Beck, T.~L., Bary, J.~S., \& McGregor, P.~J. 2010, Astrophysical Journal, 722,
  1360, \dodoi{10.1088/0004-637X/722/2/1360}

\bibitem[{Booth \& Clarke(2016)}]{Booth2016}
Booth, R.~A., \& Clarke, C.~J. 2016, Monthly Notices of the Royal Astronomical
  Society, 458, 2676, \dodoi{10.1093/mnras/stw488}

\bibitem[{Booth \& Clarke(2019)}]{Booth2019}
---. 2019, Monthly Notices of the Royal Astronomical Society, 483, 3718,
  \dodoi{10.1093/mnras/sty3340}

\bibitem[{Brandenburg(2003)}]{Brandenburg2003}
Brandenburg, A. 2003, in Advances in Non-linear Dynamos, ed. A.~Ferriz-Mas \&
  M.~Nunez (London: Taylor and Francis), 269.
\newblock \doarXiv{0109497}

\bibitem[{Cabral \& Leedom(1993)}]{Cabral1993}
Cabral, B., \& Leedom, L.~C. 1993, Proceedings of the 20th annual conference on
  Computer graphics and interactive techniques - SIGGRAPH '93, 263,
  \dodoi{10.1145/166117.166151}

\bibitem[{Carballido {et~al.}(2011)Carballido, Bai, \& Cuzzi}]{Carballido2011}
Carballido, A., Bai, X.-N., \& Cuzzi, J.~N. 2011, Monthly Notices of the Royal
  Astronomical Society, 415, 93, \dodoi{10.1111/j.1365-2966.2011.18661.x}

\bibitem[{Carrasco-Gonz{\'{a}}lez {et~al.}(2019)Carrasco-Gonz{\'{a}}lez,
  Sierra, Flock, Zhu, Henning, Chandler, Galv{\'{a}}n-Madrid, Mac{\'{i}}as,
  Anglada, Linz, Osorio, Rodr{\'{i}}guez, Testi, Torrelles, P{\'{e}}rez, \&
  Liu}]{Carrasco-Gonzalez2019}
Carrasco-Gonz{\'{a}}lez, C., Sierra, A., Flock, M., {et~al.} 2019, The
  Astrophysical Journal, 883, 71, \dodoi{10.3847/1538-4357/ab3d33}

\bibitem[{Cossins {et~al.}(2009)Cossins, Lodato, \& Clarke}]{Cossins2009}
Cossins, P., Lodato, G., \& Clarke, C.~J. 2009, Monthly Notices of the Royal
  Astronomical Society, 393, 1157, \dodoi{10.1111/j.1365-2966.2008.14275.x}

\bibitem[{Dubrulle {et~al.}(1995)Dubrulle, Morfill, \& Sterzik}]{Dubrulle1995}
Dubrulle, B., Morfill, G., \& Sterzik, M. 1995, Icarus, 114, 237.
\newblock
  \url{http://www.sciencedirect.com/science/article/pii/S0019103585710585}

\bibitem[{Dullemond \& Dominik(2004)}]{Dullemond2004}
Dullemond, C.~P., \& Dominik, C. 2004, Astronomy {\&} Astrophysics, 421, 1075,
  \dodoi{10.1051/0004-6361:20040284}

\bibitem[{Durisen {et~al.}(2007)Durisen, Boss, Mayer, Nelson, Quinn, \&
  Rice}]{Durisen2007}
Durisen, R.~H., Boss, A.~P., Mayer, L., {et~al.} 2007, in Protostars and
  Planets V, ed. B.~Reipurth, D.~Jewitt, \& K.~Keil No.~1 (Tucson: University
  of Arizona Press), 607.
\newblock \doarXiv{0603179}

\bibitem[{Flaherty {et~al.}(2015)Flaherty, Hughes, Rosenfeld, Andrews, Chiang,
  Simon, Kerzner, \& Wilner}]{Flaherty2015}
Flaherty, K.~M., Hughes, A.~M., Rosenfeld, K.~A., {et~al.} 2015, The
  Astrophysical Journal, 813, 99, \dodoi{10.1088/0004-637X/813/2/99}

\bibitem[{Flaherty {et~al.}(2017)Flaherty, Hughes, Rose, Simon, Qi, Andrews,
  Kospal, Wilner, Chiang, Armitage, \& Bai}]{Flaherty2017}
Flaherty, K.~M., Hughes, A.~M., Rose, S.~C., {et~al.} 2017, The Astrophysical
  Journal, 843, 150, \dodoi{10.3847/1538-4357/aa79f9}

\bibitem[{Flock {et~al.}(2017)Flock, Nelson, Turner, Bertrang,
  Carrasco-Gonz{\'{a}}lez, Henning, Lyra, \& Teague}]{Flock2017}
Flock, M., Nelson, R.~P., Turner, N.~J., {et~al.} 2017, The Astrophysical
  Journal, 850, 131, \dodoi{10.3847/1538-4357/aa943f}

\bibitem[{Flock {et~al.}(2020)Flock, Turner, Nelson, Lyra, Manger, \&
  Klahr}]{Flock2020}
Flock, M., Turner, N.~J., Nelson, R.~P., {et~al.} 2020, The Astrophysical
  Journal, 897, 155, \dodoi{10.3847/1538-4357/ab9641}

\bibitem[{Gammie(1996)}]{Gammie1996a}
Gammie, C.~F. 1996, The Astrophysical Journal, 457, 355.
\newblock \url{http://adsabs.harvard.edu/full/1996ApJ...457..355G}

\bibitem[{Gammie(2001)}]{Gammie2001}
---. 2001, The Astrophysical Journal, 553, 174.
\newblock \url{http://iopscience.iop.org/0004-637X/553/1/174}

\bibitem[{Gerbig {et~al.}(2020)Gerbig, Murray-Clay, Klahr, \&
  Baehr}]{Gerbig2020}
Gerbig, K., Murray-Clay, R.~A., Klahr, H., \& Baehr, H. 2020, The Astrophysical
  Journal, 895, 91, \dodoi{10.3847/1538-4357/ab8d37}

\bibitem[{Gibbons {et~al.}(2012)Gibbons, Rice, \& Mamatsashvili}]{Gibbons2012}
Gibbons, P.~G., Rice, W. K.~M., \& Mamatsashvili, G.~R. 2012, Monthly Notices
  of the Royal Astronomical Society, 426, 1444,
  \dodoi{10.1111/j.1365-2966.2012.21731.x}

\bibitem[{Hasegawa {et~al.}(2017)Hasegawa, Okuzumi, Flock, \&
  Turner}]{Hasegawa2017}
Hasegawa, Y., Okuzumi, S., Flock, M., \& Turner, N.~J. 2017, The Astrophysical
  Journal, 845, 31, \dodoi{10.3847/1538-4357/aa7d55}

\bibitem[{Hunter(2007)}]{Hunter2007}
Hunter, J.~D. 2007, Computing in Science and Engineering, 9, 90,
  \dodoi{10.1109/mcse.2007.55}

\bibitem[{Johansen \& Klahr(2005)}]{Johansen2005a}
Johansen, A., \& Klahr, H. 2005, The Astrophysical Journal, 634, 1353,
  \dodoi{10.1086/497118}

\bibitem[{Klahr \& Bodenheimer(2003)}]{Klahr2003}
Klahr, H., \& Bodenheimer, P.~H. 2003, The Astrophysical Journal, 582, 869,
  \dodoi{10.1086/344743}

\bibitem[{Klahr \& Hubbard(2014)}]{Klahr2014}
Klahr, H., \& Hubbard, A. 2014, The Astrophysical Journal, 788, 21,
  \dodoi{10.1088/0004-637X/788/1/21}

\bibitem[{Klahr \& Schreiber(2020)}]{Klahr2020}
Klahr, H., \& Schreiber, A. 2020, The Astrophysical Journal, 901, 54,
  \dodoi{10.3847/1538-4357/abac58}

\bibitem[{Lesur {et~al.}(2014)Lesur, Kunz, \& Fromang}]{Lesur2014}
Lesur, G., Kunz, M.~W., \& Fromang, S. 2014, Astronomy {\&} Astrophysics, 566,
  A56, \dodoi{10.1051/0004-6361/201423660}

\bibitem[{Liu {et~al.}(2017)Liu, Henning, Carrasco-Gonz{\'{a}}lez, Chandler,
  Linz, Birnstiel, van Boekel, P{\'{e}}rez, Flock, Testi, Rodr{\'{i}}guez, \&
  Galv{\'{a}}n-Madrid}]{Liu2017}
Liu, Y., Henning, T., Carrasco-Gonz{\'{a}}lez, C., {et~al.} 2017, Astronomy
  {\&} Astrophysics, 607, A74.
\newblock \doarXiv{1708.03238}

\bibitem[{Malygin {et~al.}(2017)Malygin, Klahr, Semenov, Henning, \&
  Dullemond}]{Malygin2017}
Malygin, M.~G., Klahr, H., Semenov, D., Henning, T., \& Dullemond, C.~P. 2017,
  Astronomy {\&} Astrophysics, 605, A30, \dodoi{10.1051/0004-6361/201629933}

\bibitem[{Manger \& Klahr(2018)}]{Manger2018}
Manger, N., \& Klahr, H. 2018, Monthly Notices of the Royal Astronomical
  Society, 480, 2125.
\newblock \doarXiv{arXiv:1807.06492v1}

\bibitem[{Nelson {et~al.}(2013)Nelson, Gressel, \& Umurhan}]{Nelson2013}
Nelson, R.~P., Gressel, O., \& Umurhan, O.~M. 2013, Monthly Notices of the
  Royal Astronomical Society, 435, 2610, \dodoi{10.1093/mnras/stt1475}

\bibitem[{Paardekooper(2012)}]{Paardekooper2012}
Paardekooper, S.-J. 2012, Monthly Notices of the Royal Astronomical Society,
  421, 3286, \dodoi{10.1111/j.1365-2966.2012.20553.x}

\bibitem[{P{\'{e}}rez \& Granger(2007)}]{Perez2007}
P{\'{e}}rez, F., \& Granger, B.~E. 2007, Computing in Science and Engineering,
  9, 21, \dodoi{10.1109/MCSE.2007.53}

\bibitem[{Pfeil \& Klahr(2019)}]{Pfeil2019}
Pfeil, T., \& Klahr, H. 2019, The Astrophysical Journal, 871, 150,
  \dodoi{10.3847/1538-4357/aaf962}

\bibitem[{Pinte {et~al.}(2016)Pinte, Dent, M{\'{e}}nard, Hales, Hill, Cortes,
  \& {De Gregorio-Monsalvo}}]{Pinte2016}
Pinte, C., Dent, W. R.~F., M{\'{e}}nard, F., {et~al.} 2016, The Astrophysical
  Journal, 816, 25, \dodoi{10.3847/0004-637X/816/1/25}

\bibitem[{Pringle(1981)}]{Pringle1981}
Pringle, J.~E. 1981, Annual Review of Astronomy and Astrophysics, 19, 137.
\newblock \url{http://adsabs.harvard.edu/full/1981ARA{\%}26A..19..137P}

\bibitem[{Ribas {et~al.}(2015)Ribas, Bouy, \& Mer{\'{i}}n}]{Ribas2015}
Ribas, {\'{A}}., Bouy, H., \& Mer{\'{i}}n, B. 2015, Astronomy and Astrophysics,
  576, A52, \dodoi{10.1051/0004-6361/201424846}

\bibitem[{Ribas {et~al.}(2020)Ribas, Espaillat, Macias, \& Sarro}]{Ribas2020}
Ribas, {\'{A}}., Espaillat, C.~C., Macias, E., \& Sarro, L. 2020, arXiv
  preprint arXiv: 2009.03323, \dodoi{10.1051/0004-6361/202038352}

\bibitem[{Rice {et~al.}(2004)Rice, Lodato, Pringle, Armitage, \&
  Bonnell}]{Rice2004}
Rice, W. K.~M., Lodato, G., Pringle, J.~E., Armitage, P.~J., \& Bonnell, I.~A.
  2004, Monthly Notices of the Royal Astronomical Society, 355, 543,
  \dodoi{10.1111/j.1365-2966.2004.08339.x}

\bibitem[{Riols \& Latter(2018)}]{Riols2018a}
Riols, A., \& Latter, H.~N. 2018, Monthly Notices of the Royal Astronomical
  Society, 476, 5115, \dodoi{10.1093/mnras/sty460}

\bibitem[{Riols {et~al.}(2017)Riols, Latter, \& Paardekooper}]{Riols2017}
Riols, A., Latter, H.~N., \& Paardekooper, S.-j. 2017, Monthly Notices of the
  Royal Astronomical Society, 471, 317, \dodoi{10.1093/mnras/stx1548}

\bibitem[{Riols \& Lesur(2018)}]{Riols2018}
Riols, A., \& Lesur, G. 2018, Astronomy and Astrophysics, 617, A117,
  \dodoi{10.1051/0004-6361/201833212}

\bibitem[{Riols {et~al.}(2020)Riols, Roux, Latter, \& Lesur}]{Riols2020}
Riols, A., Roux, B., Latter, H.~N., \& Lesur, G. 2020, Monthly Notices of the
  Royal Astronomical Society, 493, 4631, \dodoi{10.1093/mnras/staa567}

\bibitem[{Sengupta {et~al.}(2019)Sengupta, Dodson-Robinson, Hasegawa, \&
  Turner}]{Sengupta2019}
Sengupta, D., Dodson-Robinson, S.~E., Hasegawa, Y., \& Turner, N.~J. 2019, The
  Astrophysical Journal, 874, 26.
\newblock \doarXiv{arXiv:1808.03016v1}

\bibitem[{Shakura \& Sunyaev(1973)}]{Shakura1973}
Shakura, N.~I., \& Sunyaev, R.~A. 1973, Astronomy {\&} Astrophysics, 24, 337

\bibitem[{Shi \& Chiang(2014)}]{Shi2014}
Shi, J.-M., \& Chiang, E.~I. 2014, The Astrophysical Journal, 789, 34,
  \dodoi{10.1088/0004-637X/789/1/34}

\bibitem[{Stoll \& Kley(2014)}]{Stoll2014}
Stoll, M. H.~R., \& Kley, W. 2014, Astronomy {\&} Astrophysics, 572, A77.
\newblock
  \url{http://www.aanda.org/articles/aa/full{\_}html/2014/12/aa24114-14/aa24114-14.html}

\bibitem[{Stoll \& Kley(2016)}]{Stoll2016}
---. 2016, Astronomy {\&} Astrophysics, 594, A57,
  \dodoi{10.1051/0004-6361/201527716}

\bibitem[{Teague {et~al.}(2016)Teague, Guilloteau, Semenov, Henning, Dutrey,
  Pi{\'{e}}tu, Birnstiel, Chapillon, Hollenbach, \& Gorti}]{Teague2016}
Teague, R., Guilloteau, S., Semenov, D., {et~al.} 2016, Astronomy {\&}
  Astrophysics, 592, A49, \dodoi{10.1051/0004-6361/201628550}

\bibitem[{Toomre(1964)}]{Toomre1964}
Toomre, A. 1964, The Astrophysical Journal, 139, 1217.
\newblock \url{http://adsabs.harvard.edu/full/1964ApJ...139.1217T7}

\bibitem[{van~der Walt {et~al.}(2011)van~der Walt, Colbert, \&
  Varoquaux}]{vanderWalt2011}
van~der Walt, S.~J., Colbert, S.~C., \& Varoquaux, G. 2011, Computing in
  Science and Engineering, 13, 22, \dodoi{10.1109/MCSE.2011.37}

\bibitem[{Villenave {et~al.}(2020)Villenave, M{\'{e}}nard, Dent, Duchene,
  Stapelfeldt, Benisty, Boehler, van~der Plas, Pinte, Telkamp, Wolff, Flores,
  Lesur, Louvet, Riols, Dougados, Williams, \& Padgett}]{Villenave2020}
Villenave, M., M{\'{e}}nard, F., Dent, W. R.~F., {et~al.} 2020, Astronomy {\&}
  Astrophysics, 642, A164, \dodoi{10.1051/0004-6361/202038087}

\bibitem[{Virtanen {et~al.}(2020)Virtanen, Gommers, Oliphant, Haberland, Reddy,
  Cournapeau, Burovski, Peterson, Weckesser, Bright, van~der Walt, Brett,
  Wilson, Millman, Mayorov, Nelson, Jones, Kern, Larson, Carey, Polat, Feng,
  Moore, VanderPlas, Laxalde, Perktold, Cimrman, Henriksen, Quintero, Harris,
  Archibald, Ribeiro, Pedregosa, van Mulbregt, Vijaykumar, Bardelli, Rothberg,
  Hilboll, Kloeckner, Scopatz, Lee, Rokem, Woods, Fulton, Masson,
  H{\"{a}}ggstr{\"{o}}m, Fitzgerald, Nicholson, Hagen, Pasechnik, Olivetti,
  Martin, Wieser, Silva, Lenders, Wilhelm, Young, Price, Ingold, Allen, Lee,
  Audren, Probst, Dietrich, Silterra, Webber, Slavi{\v{c}}, Nothman, Buchner,
  Kulick, Sch{\"{o}}nberger, {de Miranda Cardoso}, Reimer, Harrington,
  Rodr{\'{i}}guez, Nunez-Iglesias, Kuczynski, Tritz, Thoma, Newville,
  K{\"{u}}mmerer, Bolingbroke, Tartre, Pak, Smith, Nowaczyk, Shebanov, Pavlyk,
  Brodtkorb, Lee, McGibbon, Feldbauer, Lewis, Tygier, Sievert, Vigna, Peterson,
  More, Pudlik, Oshima, Pingel, Robitaille, Spura, Jones, Cera, Leslie, Zito,
  Krauss, Upadhyay, Halchenko, \& V{\'{a}}zquez-Baeza}]{Virtanen2020}
Virtanen, P., Gommers, R., Oliphant, T.~E., {et~al.} 2020, Nature Methods, 17,
  261, \dodoi{10.1038/s41592-019-0686-2}

\bibitem[{Yang {et~al.}(2018)Yang, {Mac Low}, \& Johansen}]{Yang2018}
Yang, C.-C., {Mac Low}, M.-M., \& Johansen, A. 2018, The Astrophysical Journal,
  868, 27, \dodoi{10.3847/1538-4357/aae7d4}

\bibitem[{Youdin \& Lithwick(2007)}]{Youdin2007a}
Youdin, A.~N., \& Lithwick, Y. 2007, Icarus, 192, 588,
  \dodoi{10.1016/j.icarus.2007.07.012}

\bibitem[{Zhu {et~al.}(2015)Zhu, Dong, Stone, \& Rafikov}]{Zhu2015}
Zhu, Z., Dong, R., Stone, J.~M., \& Rafikov, R.~R. 2015, The Astrophysical
  Journal, 813, 88.
\newblock \doarXiv{1507.03599}

\bibitem[{Zhu {et~al.}(2012)Zhu, Hartmann, Nelson, \& Gammie}]{Zhu2012}
Zhu, Z., Hartmann, L.~W., Nelson, R.~P., \& Gammie, C.~F. 2012, The
  Astrophysical Journal, 746, 110, \dodoi{10.1088/0004-637X/746/1/110}

\end{thebibliography}

\appendix

\section{Diffusion Constant}
\label{sec:appendixdiffusionconstant}

The diffusion constant derived in the text uses a simpler relation between particle scale height $H_{p}$ and particle size $t_{s}$ based on an assumption of no vertical stratification. The simulations presented are clearly stratified, but the relation still holds well for the intermediate sizes and is easier to interpret from Figure \ref{fig:diffusionparameter}. Nevertheless, we find it important to validate the result with a more rigorous relation and we provide an alternate derivation of the diffusion constant based on \citet{Dubrulle1995} which allows for a fit which includes $\mathrm{St}=0.01$
\begin{equation} \label{eq:dubrullediffusion}
H_{d} = H_{g} \sqrt{\frac{1}{(t_{s}\Omega^{2}H_{g}^{2})/\mathcal{D}_{d,z}+1}}.
\end{equation}
We rewrite this equation as
\begin{equation} \label{eq:altdubrullediffusion}
\eta = \mathcal{D}_{d,z}^{-1}t_{s}\,,
\end{equation}
where $\eta = H_{d}^{-2} - H_{g}^{-2}$, to evaluate a dimensionless diffusion constant $\mathcal{D}_{d,z}^{-1}$ with a linear fit comparable to Figure \ref{fig:diffusionparameter}. The diffusion constants derived in this way are (in the same descending order as in Table \ref{tab:turbresults}) $\mathcal{D}_{d,z}=$ [$1.9\times10^{-3}$, $1.6\times10^{-3}$, $1.0\times10^{-3}$, $8\times10^{-3}$, $3.0\times10^{-3}$, $2.9\times10^{-3}$, $1.3\times10^{-3}$, $4.2\times10^{-3}$].

\begin{figure}
\centering
\includegraphics[width=0.48\textwidth]{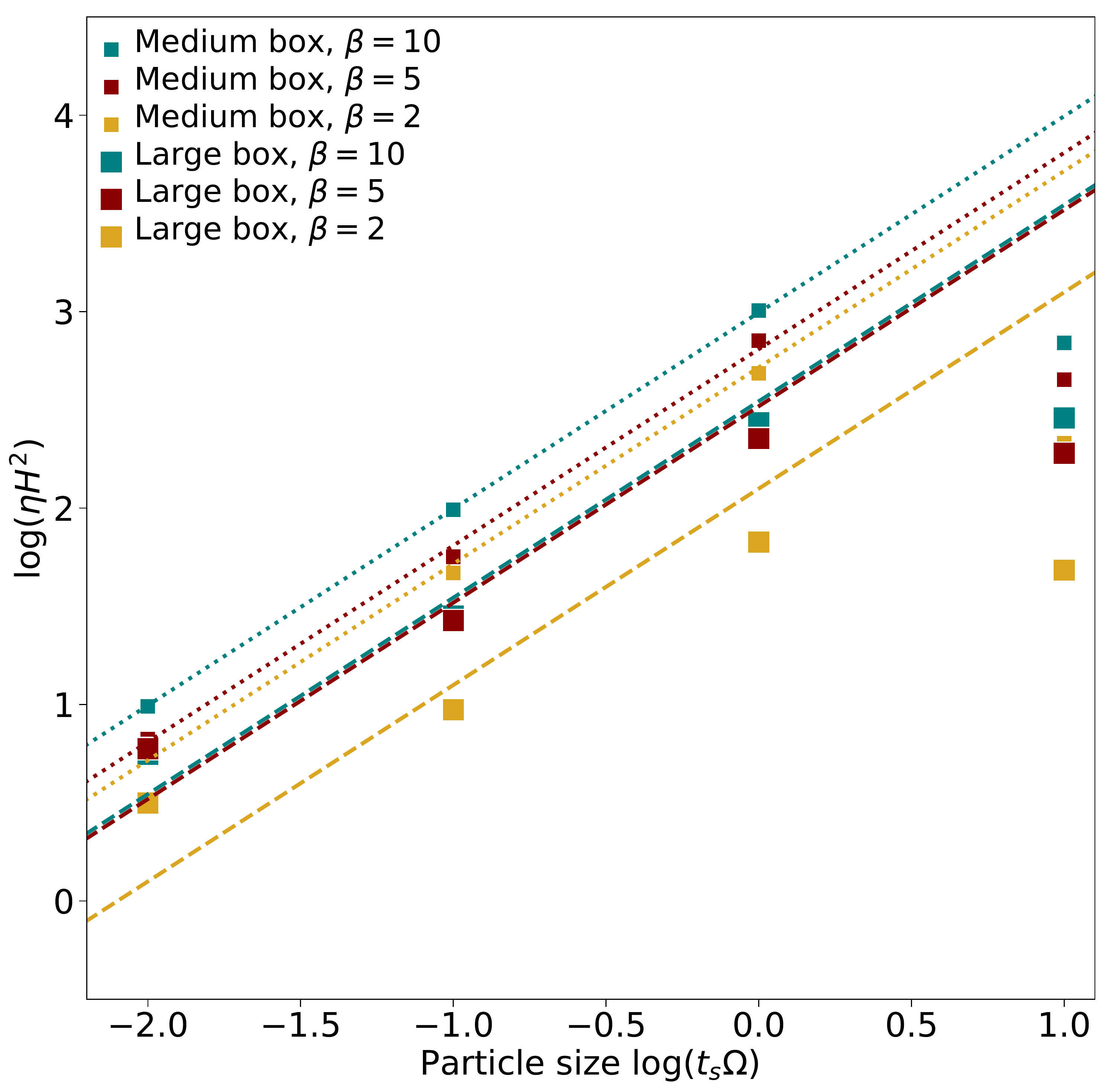}%
\includegraphics[width=0.48\textwidth]{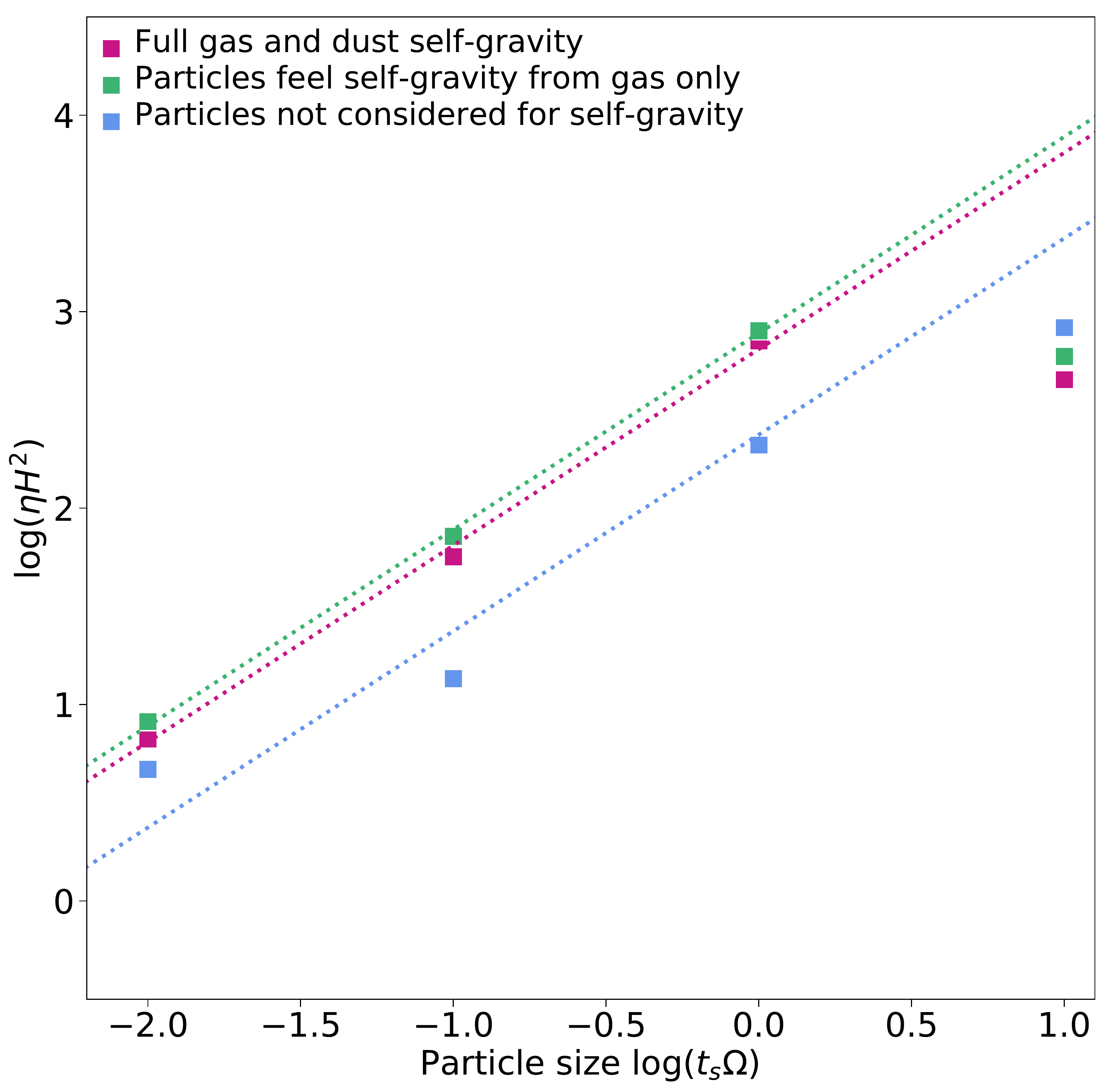}
\caption{Left: Approximating $\mathcal{D}_{d,z}$ for the six primary simulations of this paper, using Equation \ref{eq:altdubrullediffusion}; compare with Figure \ref{fig:diffusionparameter}. Right: $\mathcal{D}_{d,z}$ for the simulations which investigate the self-gravity implementation as in the left panel of Figure \ref{fig:selfgravitycomparison}.}
\label{fig:dubrullediffusionparameter}
\end{figure}

\end{document}